\documentclass[12pt,preprint]{aastex}

\newcommand{\myemail}{lauren@astro.ubc.ca}
\newcommand{\etal}{et al.\@}
\newcommand{\sersic}{S\'{e}rsic}

\newcommand{\nfit}{$n_{\hbox{\small{fit}}}$}
\newcommand{\nmodel}{$n_{\hbox{\small{model}}}$}
\newcommand{\hfit}{$h_{\hbox{\small{fit}}}$}
\newcommand{\hmodel}{$h_{\hbox{\small{model}}}$}

\newcommand{\chisqrin}{$\chi^{2}_{in}$}
\newcommand{\chisqrgl}{$\chi^{2}_{gl}$}
\newcommand{\chisqr}{$\chi^2$}
\newcommand{\eg}{e.g.}
\newcommand{\ie}{i.e.}
\newcommand{\hunit}{km~sec$^{\hbox{\scriptsize -1}}$~Mpc$^{\hbox{\scriptsize -1}}$}
\newcommand{\hub}{H$_{\hbox{\scriptsize 0}}$}
\newcommand{\magarc}{\ifmmode {{{{\rm mag}~{\rm arcsec}}^{-2}}}
             \else {{{mag}$~${arcsec}$^{-2}$}}
             \fi}
\newcommand{\kms}{\ifmmode\,{\rm km}\,{\rm s}^{-1}\else km$\,$s$^{-1}$\fi}

\def \lapprx {$_ <\atop{^\sim}$}

\shorttitle{Structure of Disk Dominated Galaxies.I.}
\shortauthors{MacArthur, Courteau, \& Holtzman}

\received{2002 March 15}
\begin{document}

\title{Structure of Disk Dominated Galaxies I. \ Bulge/Disk Parameters,
 Simulations, and Secular Evolution}

\author{Lauren A.~MacArthur and St\'{e}phane Courteau\altaffilmark{1}}

\affil{Department of Physics \& Astronomy, University of British Columbia,
    6224 Agricultural Road, Vancouver, BC V6T 1Z1}
\email{\myemail, courteau@astro.ubc.ca}

\and

\author{Jon A.~Holtzman\altaffilmark{1}}

\affil{Department of Astronomy, New Mexico State University,
     Box 30001, Department 45000, Las Cruces, NM 88003-8001}
\email{holtz@astro.nmsu.edu}

\altaffiltext{1}{Visiting Astronomer at Lowell and Kitt Peak National
   Observatory. KPNO is operated by AURA, Inc. under
   cooperative agreement with the National Science Foundation.}

\begin{abstract}
A robust analysis of galaxy structural parameters, based on the modeling
of bulge and disk brightnesses in the BVRH bandpasses, is presented for
121 face-on and moderately inclined late-type spirals.
Each surface brightness (SB) profile is decomposed into a sum of a
generalized \sersic\ bulge and an exponential disk.
The reliability and limitations of our bulge-to-disk (B/D) decompositions
are tested with extensive simulations of galaxy brightness profiles (1D) and
images (2D)\@. We have used repeat observations to test the consistency
of our decompositions.  The average systematic model errors are \lapprx20\%
and \lapprx5\% for the bulge and disk components, respectively.
The final set of galaxy parameters is studied for
variations and correlations in the context of profile type differences
and wavelength dependences.

Galaxy types are divided into three classes according to their SB profile
shapes; Freeman Type-I and Type-II,
and a third ``Transition'' class for galaxies whose profiles change
from Type-II in the optical to Type-I in the infrared.  Roughly 43\%,
44\%, and 13\% of Type I, II, and Transition galaxies respectively
comprise our sample.  Only Type-I galaxies, with their fully exponential
disks, are adequately modeled by our 2-component decompositions and our
main results focus on these profiles.  We discuss possible interpretations
of Freeman Type-II profiles.

The \sersic\ bulge shape parameter for nearby Type-I late-type spirals shows
a range between $n=$0.1--2 but, on average, the underlying surface density
profile for the bulge and disk of these galaxies is adequately described
by a double-exponential distribution.
The distribution of disk scale lengths shows a decreasing trend with
increasing wavelength, consistent with a higher concentration of old
stars or dust (or both) in the central regions relative to the outer disk.
We confirm a coupling between the bulge and disk with a scale length
ratio $\langle r_e/h \rangle = 0.22 \pm 0.09$, or $\langle h_{bulge}/h_{disk}
\rangle = 0.13 \pm 0.06$ for late-type spirals, in agreement with recent
N-body simulations of disk formation.  This ratio increases from $\sim0.2$
for late-type spirals to $\sim0.24$ for earlier types.
These observations are consistent with bulges of late-type spiral galaxies
being more deeply embedded in their host disk than earlier-type bulges, as 
discussed by Graham (2001). Bulges and disks
can thus preserve a nearly constant $r_e/h$ but show a great range of
surface brightness for any given effective radius.  
The similar scaling relation for early and
late-type spirals suggests comparable formation and/or evolution scenarios
for disk galaxies of all Hubble types. In the spirit of Courteau, de~Jong,
\& Broeils (1996)
but using our new, more extensive data base, we interpret this result as
further evidence for regulated bulge formation by redistribution of disk
material to the galaxy center, in agreement with models of secular evolution
of the disk.

\end{abstract}

\keywords{galaxies: spiral---galaxies: photometry---galaxies: structure
---galaxies: formation---galaxies: simulations}

\section{Introduction}\label{sec:intro}

Stellar density distributions provide important constraints
for bulge and disk formation models.
Historically, astronomers have embraced the $r^{1/4}$
brightness ``law'' \citep{deVauc48} and exponential brightness
profile\footnote{The exponential nature of galaxy disk profiles
emerges naturally in analytical models of disk formation
(e.g.\@ Lin \& Pringle 1987; Dalcanton, Spergel, \& Summers 1997;
 Ferguson \& Clarke 2001).} (de Vaucouleurs 1959a; Freeman 1970)
to model the light distribution of the galaxy bulge and
disk, respectively\footnote{It is important to remind ourselves from
the onset that bulge-to-disk decompositions,
and inward extrapolations of the disk into the central bulge and/or
bar, may have no physical (or dynamical) basis. They provide a
convenient description of the light distribution of a galaxy's components
that are otherwise dynamically coupled.  The effective integrals of
motions are likely similar for all the co-spatial components, though
kinematically distinct bulges (counter-rotating nuclei) are known
to exist.}. Departures from the standard de~Vaucouleurs
profile in the {\it inner} light distribution of early- and late-type
spirals have however been demonstrated in a number of early studies
\citep{deVauc59, vanHout61, Burstein79}, including the Milky Way
\citep{Kent91}.  \citet{AndSan94}, \citet{deJong96a}, Courteau, de~Jong,
\& Broeils (1996),
and \cite{Carollo99} later used small samples of high-quality surface
brightness (SB) profiles to establish the exponential profile as a better
match to {\it late}-type disk bulges; thus SB profiles
of most late-type spirals are best modeled by a
double-exponential fit to the bulge and disk.

A broader analysis suggests a range of bulge shapes from early- to late-type
spirals \citep{AndPelBal95, deJong96a, CourdeJBro96, Graham01}.
Most of these analyses rely on the modeling of a generalized surface density
function such as that proposed by \citet{Sersic68};
\begin{equation}
\label{eq:sersic}
I(r)=I_{0}\,{\exp\left\{-{\left({r\over{r_{0}}}\right)^{\rm{1/n}}}\right\}}
\end{equation}
or, in magnitudes,
\begin{equation}
\label{eq:sersicmag}
\mu(r)=\mu_{0}+
     2.5\log(e){\left\{\left({r\over{r_{0}}}\right)^{\rm{1/n}}\right\}}.
\end{equation}
where $\mu_0$ ($I_{0}$) is the central surface brightness (intensity),
$r_{0}$ is a scaling radius, and the exponent $1/n$ is a shape parameter that
describes the amount of curvature in the profile.  For $n=1$ or 4 one recovers
a pure exponential or the de~Vaucouleurs $r^{1/4}$ profile respectively.

Collectively, the works above suggest that the bulge shape parameter $n$
correlates with absolute luminosity and half-light radius, such that bigger,
brighter systems have larger values of $n$.  This result was extended to
brightest cluster galaxies by \citet{Graham96}.
\citet{CourdeJBro96} also demonstrated a tight correlation between the bulge
and disk exponential scale lengths, for all spiral types, with
$h_b/h_d = 0.1\pm0.05$ (where $h=r_{0}$ and $n=1$ in Eq.~\ref{eq:sersic}).
The exponential nature of late-type galaxy bulges and the correlation between
bulge and disk scale lengths was interpreted by \citet{CourdeJBro96}
as evidence for regulated bulge formation by redistribution of disk
material to the galaxy center by a bar-like perturbation.
We will return to this important constraint for secular evolution models
in \S~\ref{subsec:sec_ev}.

This study focuses on the development of a reliable set of observables
and constraints for structure formation models.  An important goal is to
measure the range of the \sersic\ $n$ parameter for virialized
disk systems. The analyses
described above are reproduced and expanded upon with the largest multi-band
survey of its kind to date and a clearer understanding of model limitations
than previously attained.
We aim to characterize and quantify the intrinsic structural properties
of the bulge and disk and the extent of their variation with wavelength.
These characterizations are made through reliable modeling
of bulge and disk parameters from SB profile decompositions.
Multi-wavelength information also provides insight about structural
variations within and among galaxies due to dust and stellar population
effects.  While some of these issues have been addressed before, there remains
a number of significant measurement uncertainties and technical limitations
which we now investigate thoroughly.

This paper is organized as follows: a brief description of the database
is given in \S~\ref{sec:data} and in \S~\ref{sec:simulations} we discuss our
B/D decomposition
algorithms (1D and 2D) and the simulations to test the reliability
of our technique.
For the readers interested mostly in final profile decompositions and
results, a summary of the simulation results and guidelines is given
in \S~\ref{sec:sim_summ}.
Actual B/D decompositions of galaxy SB profiles are presented
in \S~\ref{sec:decomps}, followed by a discussion
and interpretation of the results in terms of secular evolution models in
\S~\ref{sec:results}.
A discussion on the nature of Freeman Type-II profiles is also presented
in \S~\ref{sec:results}.
We conclude with future directions in \S~\ref{sec:summary}.  Two
appendices present (A) a discussion of the functional form for the
\sersic\ coefficient $b_n$, and (B) decomposition results for our
Type-I profiles.

\section{The Data}\label{sec:data}

Our structural analysis of galaxy luminosity profiles is based on the
catalog of multi-band images of late-type spiral galaxies by Courteau,
Holtzman, \& MacArthur (2002; hereafter Paper II).  It consists of over 1000
deep B, V, R, and H images of 322 nearby bright late-type spiral galaxies.
The data were collected between 1992 and 1996 at Lowell Observatory
and Kitt Peak National Observatory (KPNO).  A full description of the sample
selection, observations, and reductions is presented in Paper II\@.  A summary
is given below.

The galaxy sample was selected from the Uppsala General Catalogue
(UGC, Nilson 1973\nocite{Nilson73}) with the following criteria:
\begin{itemize}
\item Predominently late Hubble types
\vspace{-0.3cm}
\item Zwicky magnitude $m_B \leq$ 15.5
\vspace{-0.3cm}
\item Blue Galactic extinction $A_B=4 \times E(B-V) \leq$ 0\fm5
     \ \ (Burstein \& Heiles 1984\nocite{BursHeil84})
\vspace{-0.3cm}
\item Inclination bins covering face-on ($i\leq$ 6\degr), intermediate
     (50$^\circ< i <$ 60$^\circ$), and edge-on ($i\geq$ 78\degr) projections
\vspace{-0.3cm}
\item Blue major axis $\leq 3\arcmin$ .
\end{itemize}

This catalog is not complete in any sense of the (much abused) term.
The diameter limit was constrained primarily by the field of view
of the infrared cameras in use at KPNO (IRIM and COB) and Lowell
Observatory (OSIRIS) in 1992--1996 and
the requirement for blank areas in the field of view for sky
subtraction.  Additionally, peculiar and interacting galaxies (e.g.\@ no
visible tidal tails) were excluded to ensure that the sample consisted only
of isolated disk dominated galaxies.  Barred galaxies, as classified in
the UGC, were not excluded {\it per se} but only a handful were observed.
For the present analysis, we use a sub-sample of 121
galaxies with face-on and intermediate inclinations only, for a
total of 523 images\footnote{Radial brightness profiles cannot be measured
for fully edge-on galaxies.  Our images of edge-on systems will be used
in a forthcoming analysis
of stellar and dust scale heights and truncation radii in spiral disks.}.
The distribution of Hubble types in our reduced sample is: 2~Sab, 26~Sb,
19~Sbc, 38~Sc, 25~Scd, 11~Sd.

All distances are corrected to the reference frame of the
Local Standard of Rest \citep{CourvdB99}, and we use \hub~=~70 \hunit.
The survey effective depth is $\langle cz \rangle \sim 5500 \kms$ or
$80$ Mpc.

\subsection{Observations and Basic Reductions}

All optical BVR images were obtained from 1992 to 1994 at Lowell Observatory
with a TI 800$\times$800 chip (scale = 0\farcs5/pix) on the Perkins
72\arcsec\ telescope.
The infrared H-band images were acquired from 1993 to 1995 at KPNO with the
2-meter and 4-meter telescopes equipped
with either a HgCdTe (IRIM) or an InSb (COB) 256$\times$256 array
(1\farcs09/pix and 0\farcs5/pix respectively), and from 1995
to 1996 with the OSIRIS imager (1\farcs 49/pix) mounted on the
Lowell 72\arcsec\ telescope.  The exposure times were: 300s in R,
400s in V, 1500s in B, and on-target integration of 1200s in H\@.
\citet{Landolt92} standards covering a wide range of airmasses
and colors were observed each night at Lowell Observatory, giving
a photometric accuracy of $\sim$2\% for the optical passbands.
UKIRT standards (Guarnieri, Dixon, \& Longmore 1991\nocite{Guarnieri91})
observed each night yielded H-band photometric calibrations good to
$\sim$3\%.  Stars and defects were edited from the images prior to further
analysis.

The typical seeing full-width at half-maximum (FWHM) at Lowell and KPNO
was 2\farcs0 with typical standard deviations of $\sim$20\% (optical) and
$\sim$35\% (IR) per image.  These measurements were computed as the mean of
the FWHMs of all non-saturated stars measured automatically on each image
frame; typically 5 to 40 measurements per frame were used.

We measure mean sky levels (for a 5--6 day old moon) of $B=21.9 \pm 0.8,
V=21.2 \pm 0.5, R=20.6 \pm 0.5$, and $H=14.1 \pm 1.2$ \magarc. Typical
systematic errors in the sky measurement, computed from 4 or 5 sky boxes
suitably located between the galaxy and the edge of the frame, are $0.5-1.0\%$
in the optical and $0.005-0.01\%$ in the IR\@.

Azimuthally-averaged SB profiles were extracted for all
the galaxies using ellipse fitting with a fixed center.  To ensure a
homogeneous computation of structural parameters and
color gradients, we use the
isophotal maps from the R-band to determine the SB profiles in BVH\@.
Even though dust effects can still play a role at 7000 \AA, the R-band
was adopted for our isophotal templates as it has the most stable sky and
deepest profiles.
We allowed a variable position angle and ellipticity at each
isophote, but a comparison with SB profiles extracted using concentric
isophotal fits demonstrated that our results do not depend on the fitting
technique.
Further information about profile extraction and
CCD surface photometry can be found in Courteau (1996a) and Paper~II\@.
We trace SB profiles to $\sim$26~\magarc
in optical bands and $\sim$22~\magarc at H-band.  These levels correspond
to a surface brightness error of $\sim$0.12~\magarc.

\subsection{Surface Brightness Corrections}\label{sec:sbcorr}

The observed surface brightness of a galaxy can change when viewed at
different inclination angles, depending on the distribution of a galaxy's
interstellar medium and its opacity.  Surface brightnesses are also
affected by Galactic foreground extinction and redshift dimming.
We account for the latter effects but defer any treatment of internal
extinction, which vary greatly from author to author, to Paper II\@.
Our conclusions do not depend on the exact values of the central
and effective surface brightnesses of galaxies.

We correct for Galactic foreground extinction using the reddening
values, $A_{\lambda}$, of Schlegel \etal\ (1998) and assuming an R$_V$ = 3.1
extinction curve (\eg\ Cardelli \etal\ 1989\nocite{Cardelli89}),
\begin{equation}
\mu^{\lambda}_{c,Gal} = \mu^{\lambda}_{obs} - A_{\lambda}.
\end{equation}

We correct surface brightnesses for the $(1+z)^3$ cosmological
redshift dimming ({\it per unit frequency interval}) as
\begin{equation}
\mu^{\lambda}_{c,z} = \mu^{\lambda}_{obs} - 7.5\log(1+z).
\end{equation}

The final correction to the observed surface brightnesses is thus,
\begin{equation}
\mu^{\lambda}_c = \mu^{\lambda}_{obs} - A_{\lambda}  - 7.5\log(1+z).
\end{equation}

\begin{figure}
\plotone{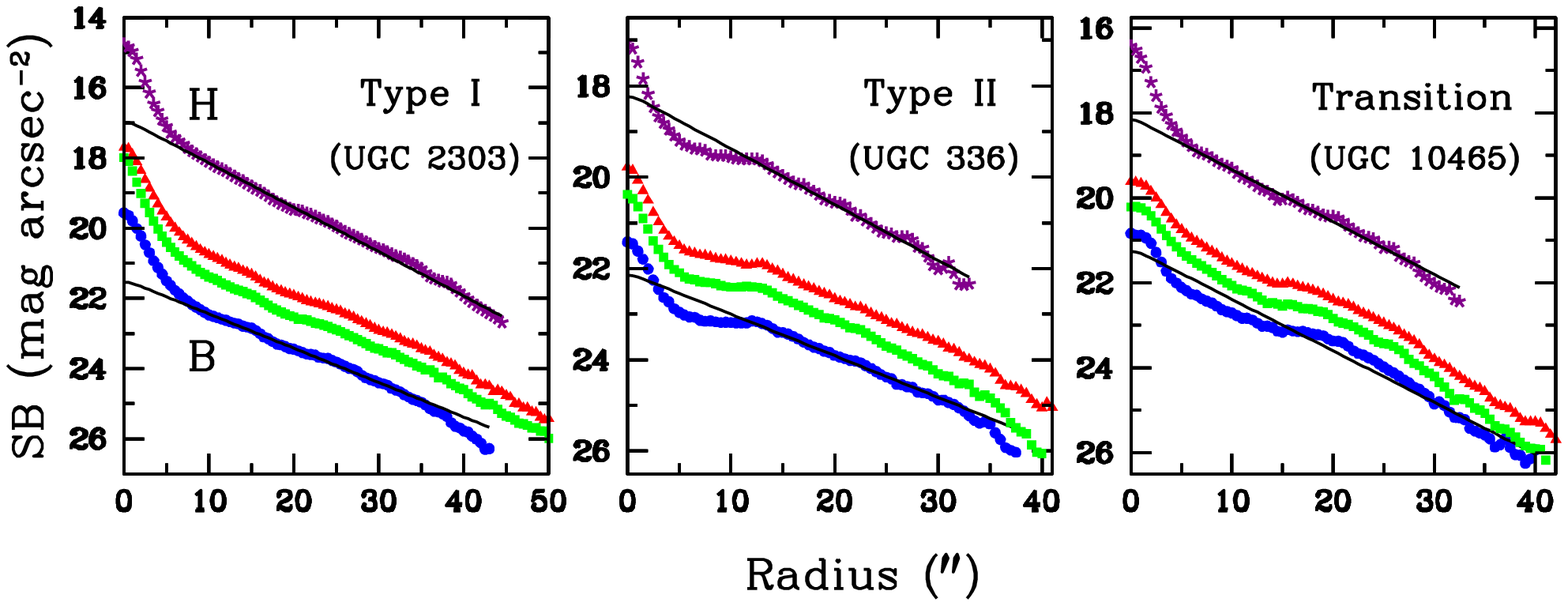}
    \caption{Examples of Type-I (left), Type-II (middle), and ``Transition''
    (right) SB profiles.  Blue circles, green squares, red triangles, and 
     purple asterisks are for B, V, R, and H-band respectively.  The 
     solid black lines plotted on the B-band and H-band profiles are fits to 
     the outer exponential disk profile.
   \label{fig:types}}
\end{figure}

Examples of the types and extent of the SB profiles
for typical late-type spiral galaxies in our sample are shown in
Fig.~\ref{fig:types}.  For Type-I disks \citep{Freeman70},
the inner profile always lies above the surface brightness of the
inward extrapolation of the outer disk, whereas Type-II systems
have a portion of their brightness profiles lying below the
inward disk extrapolation.
We define a {\it Transition} case for luminosity profiles that change from
Type-II at optical wavelengths to Type-I in the infrared.
Many galaxies classified as Type-II show a weakening of the inner
profile dip at longer wavelengths and, in this sense there is no clear
distinction between the Type-II and Transition galaxies.  Likely
interpretations for Type-II profiles are discussed in \S~ref{subsec:typeII}.

\section{Simulations of Bulge-to-Disk Decompositions}\label{sec:simulations}

In order to measure galaxy structural parameters, we have developed two
independent algorithms to decompose the galaxy 1D and 2D light distributions
into bulge and disk components.  These
programs allow for a generalized \sersic\ bulge, an exponential disk,
and a central bar for 2D images.   There
are several issues involved with accurate decompositions, particularly
with the measurement of bulge parameters, including; the sensitivity of final
results to starting guesses, effects of statistical and systematic errors in
sky brightness and seeing estimates, choice of fit baseline, etc.  We explore
these in great detail below using both 1D and 2D analyses to determine
the robustness of our codes and the reliability of our final solutions.
Because projected surface brightness profiles contain fewer data points than
full 2D images, we can create 1D
simulations faster than 2D models.
Thus our most extensive tests rely on 1D simulations, which are shown
to be fully consistent with 2D simulations when considering axisymmetric
features.

\subsection{1D and 2D Algorithms}\label{subsec:algorithm}

Our brightness profile (1D) bulge-to-disk (B/D) decomposition algorithm was
initially
developed by \citet{BroCou97} and subsequently improved by LM\@.  This
program reduces 1D projected galaxy luminosity profiles
into bulge and disk components simultaneously using a non-linear
Levenberg-Marquardt least-squares (NLLS; see \S~15.5 in \citet{NR})
fit to the logarithmic intensities (\ie\ magnitude units).
Random SB errors are accounted for in the (data$-$model) minimization, whereas
systematic errors such as uncertainties in the sky background and
determination of the image mean PSF are accounted for separately in a series
of experiments designed to calibrate their effects.
Seeing effects in our model galaxies are accounted for by convolving
the theoretical bulge-disk surface brightness profiles and images with
a radially symmetric Gaussian Point Spread Function (PSF), of the form
\begin{equation}
\label{eq:seeing}
I_{s}(r)=\sigma^{-2}e^{-r^{2}/2\sigma^{2}}\int_{0}^{\infty}
 I_{total}(x)I_{0}(xr/\sigma^{2})e^{-x^{2}/2\sigma^{2}}x\,dx
\end{equation}
where $I_{total}(x)$ is the intrinsic surface brightness profile, $\sigma$ is
the dispersion of the Gaussian PSF
and $I_0$ is the zero-order modified Bessel function of the first kind
(see also \citet{Trujillo01} for a study of the Moffat PSF\@.)

\bigskip

The 2D decomposition program is based on the same NLLS technique as above
but uses the full 2D image in intensity units instead of a logarithmic
radial surface brightness profile.  While computationally
more intensive than its 1D analogue, the 2D decomposition should yield
more physically meaningful results since the azimuthal
information is lost in 1D profiles.  \citet{ByuFre95},
\citet{deJong96a}, and \citet{Simard02} have discussed the merits of the
2D approach, such as
greater ability to recover true parameters (based on simulations), and the
potential to model non-axisymmetric features such as bars, rings, and spiral
arms.  The need for the implementation and testing of a robust 2D B/D
decomposition package is thus obvious, but we find that 1D
decompositions compare favorably for reliability and predictive
power provided high S/N 1D radial profiles are used.  Note that neither
1D nor 2D decompositions are impervious to dust extinction effects.
Extinction effects are lessened at H-band, but can still be significant
in disk bulges and spiral arms.  A proper recovery of the true stellar
density profile would require a full 3D radiation transfer treatment,
and such an analysis is beyond the scope of this work.

\subsection{Methodology}\label{sec:method}

A fundamental aspect of profile decompositions is the choice of fitting
functions.  The disk light is modeled with the usual exponential function,

\begin{equation}
\label{eq:exp}
I_d(r)=I_0\,\exp\left\{- {r\over{h}} \right\}
\end{equation}
or, in magnitudes,
\begin{equation}
\label{eq:expmag}
\mu_d(r)=\mu_0 + 2.5\log(e){\left\{{r\over{h}}\right\}}
\end{equation}
where $\mu_0 \equiv -2.5\log I_0$ and $h$ are the disk central surface
brightness (CSB) and scale length respectively, and $r$ is the galactocentric
radius measured along the major axis.  In the 2D decompositions,
the computation of the radius at each pixel requires two additional
parameters: the position angle (PA) of the disk major axis on the sky and the
disk ellipticity, $\varepsilon = 1 - b/a$, where $a$ and $b$ are the
major and minor axes of the disk respectively).
To test for the shape of the bulge luminosity profiles we adopt the
generalized formulation of \sersic\ (Eqs.~\ref{eq:sersic} \&
\ref{eq:sersicmag}).

It has become customary to express the disk parameters in terms of scale
length and CSB ($h$ and $\mu_0$), while the bulge parameters are expressed
in terms of {\it effective} parameters ($r_e$ and $\mu_e$).  We adopt this
formalism, thus parameters with subscript $e$ refer to the bulge.
Eq.~\ref{eq:sersic} can be re-written as:
\begin{equation}
\label{eq:sersicbn}
I_b(r)=I_{e}{\exp{\left\{-b_{n}{\left[\left({r\over{r_{e}}}\right)
^{\rm{1/n}}-1\right]}\right\}}}
\end{equation}
where the effective radius, $r_e$, encloses half the total extrapolated
luminosity\footnote{For a pure exponential disk, $r_e = 1.678 h$.}.
$I_e$ is the intensity at this radius and $b_n$ is chosen to ensure that
\begin{equation}
\int_{0}^{\infty}I_{b}(r)\,2\pi r\,dr = 2\int_{0}^{r_{e}}I_{b}(r)\,2\pi r\,dr.
\label{eq:bnderv}
\end{equation}
In magnitudes Eq.~\ref{eq:sersicbn} translates to
\begin{equation}
\label{eq:sersicbnmag}
\mu_{b}(r)=\mu_{e} +
    2.5\log(e)\,b_{n}{\left[\left({r\over{r_{e}}}\right)^{1/n} - 1\right]}
\end{equation}
where $\mu_e$ is the effective surface brightness.

It is trivial to convert from Eq.~\ref{eq:sersicmag} to
Eq.~\ref{eq:sersicbnmag} by noting that
\begin{eqnarray}
r_e &=& (b_n)^{n} r_0 \label{eq:r0tore}\\
\mu_e &=& \mu_0 + 2.5\log(e)\,b_n.
\end{eqnarray}

Eq.~\ref{eq:bnderv} implies that
\begin{equation}
\label{eq:bnexact}
\Gamma(2n) = 2\gamma(2n,b_n)
\end{equation}
where $\Gamma(a)$ is the gamma function and $\gamma(a,x)$ is the
incomplete gamma function.  Unfortunately, Eq.~\ref{eq:bnexact} cannot be
solved analytically for $b_n$.  Various numerical approximations have been
been given in the literature (Caon \etal\ 1993; Graham \& Prieto  1999;
Ciotti \& Bertin 1999; Khosroshahi, Wadadekar, \& Kembhavi 2000;
M{\" o}llenhoff \& Heidt 2001).
One often encounters the approximation $b_n \approx 2n - 0.32$, valid
supposedly for all values of $n$ (sic).  \citet{KhoWadKem00} contend that
this approximation is accurate to one part in $10^5$, with a range of validity
on $n$ unspecified.  However, because the gamma function diverges near the
origin, most utilized approximations are inaccurate for values of the \sersic\
exponent $n \leq 1.$  Differences between numerical solutions for $b_n$
(Eq.~\ref{eq:bnexact}) and commonly adopted approximations can
yield brightness differences greater than 0.1~\magarc\ for $n \la 1$.
As we wish to test for bulges with \sersic\ $n$ parameter smaller than 1,
we have adopted the asymptotic expansion of \citet{CiottiBer99}
to O$(n^{-5})$ for $n>0.36$.  For $n\leq 0.36$ this solution diverges and
instead we use a polynomial expression ($4^{th}$ order)
accurate to one part in $10^{3}$.  We compare different numerical solutions
for $b_n$ (Fig.~\ref{fig:bncomp}) and present our adopted functional form
in Appendix A\@.

\begin{figure}
\plotone{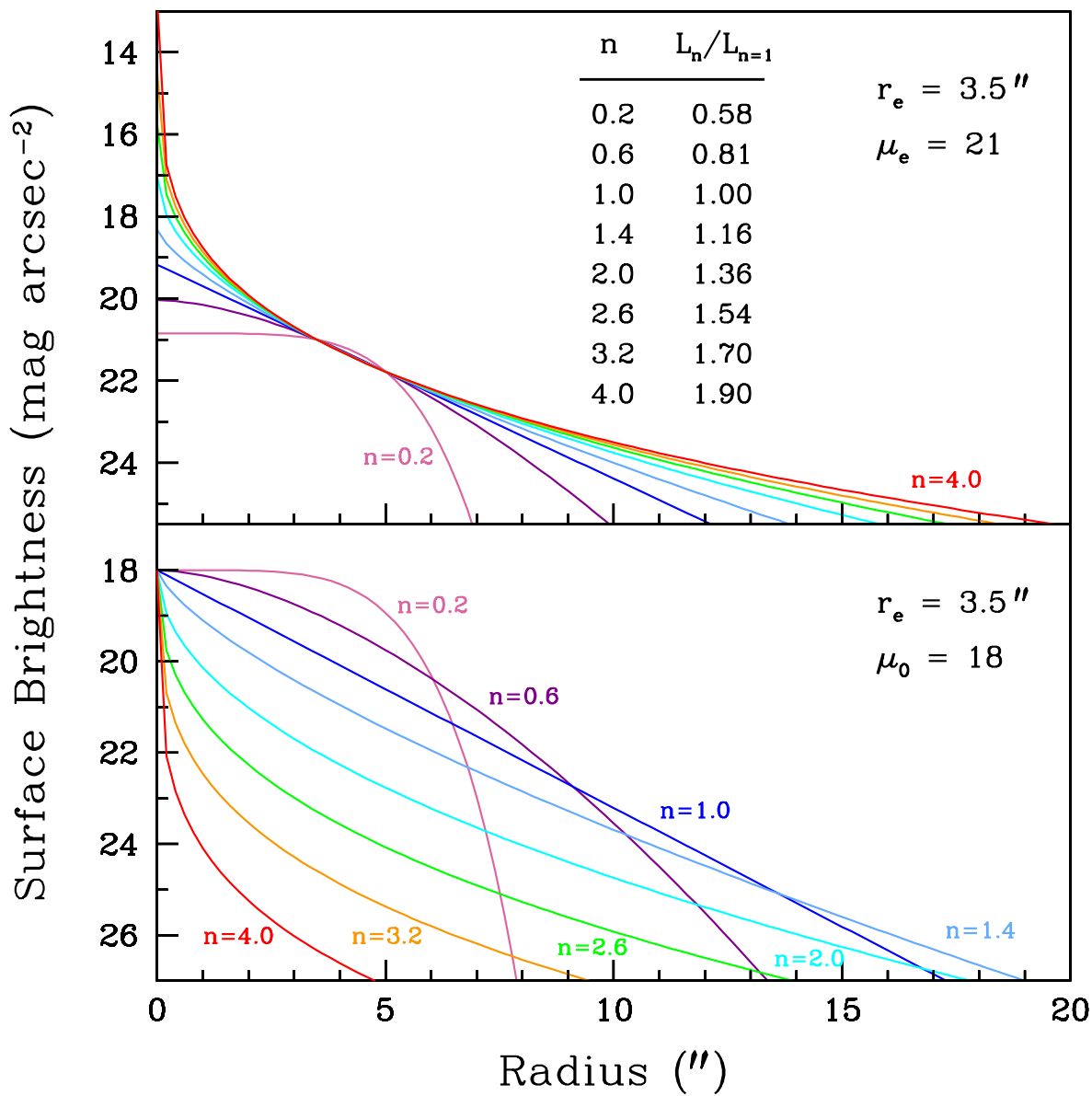}
   \caption{\sersic\ $n$
    profiles for different values of $n$.  The top panel shows profiles with
    $\mu_{e}=21$ \magarc and $r_{e}$ = 3\farcs5 for values of $n$ in the
    range $0.2<n<4$.  The table lists the relative light contributions of the
    different profiles normalized to the $n=1$ case.  The bottom panel shows
    the same profiles except for a constant CSB of
    $\mu_{0}=18$ \magarc. \label{fig:sersicn}}
\end{figure}

An illustration of profile shapes for different values of the \sersic\ $n$
parameter is shown in Fig.~\ref{fig:sersicn}.  The top panel shows profiles
with $\mu_{e}=21$~\magarc and $r_{e}$ = 3\farcs5 for values of $n$ in the
range $0.2<n<4$.  The bottom panel shows the same profiles but for a
constant CSB of $\mu_{0}=18$~\magarc.  For $n<1$ the
profiles are shallow at small radii ($\la r_{e}$) and fall off rapidly with
increasing radius.  Conversely, profiles with $n>1$ are steep at
small radii ($\ll r_e$), but level off as $r$ increases.
Given the large differences in the profile shapes above and below $n=1$
(exponential case),
one might expect different physical mechanisms (formation, transport,
dynamics, interactions) to be at work for systems whose light profiles
have very different $n$ values.  Additionally, for the small bulges of
late-type galaxies, poor seeing could conceivably smear the image such
that an intrinsically $n>1$ bulge could be mistaken for an $n<1$ structure.

Of potential relevance to the study of galaxy structure is the relative
light fraction contributed by the bulge and disk.  This is expressed in terms
of a bulge-to-disk luminosity ratio, $B/D$, derived by integrating the bulge
and disk luminosity profiles to infinity.  For a face-on \sersic\ profile
the total extrapolated luminosity is given by

\begin{equation}
\label{eq:totalbulge}
L_{b}=\int_{0}^{\infty}I_{b}(r)2\pi r\,dr={{2\pi I_{e}r_{e}^{2}e^{b_{n}
\,}n\Gamma(2n)}\over{b_{n}^{2n}}}
\end{equation}
and for a face-on exponential disk
\begin{equation}
\label{eq:totaldisk}
L_{d}=\int_{0}^{\infty}I_{d}(r)2\pi r\,dr=2\pi I_{0}h^{2}
\end{equation}
giving a bulge-to-disk light ratio of
\begin{equation}
\label{eq:bdrat}
B/D={{e^{b_{n}\,}n\Gamma(2n)}\over{b_{n}^{2n}}}
\left({r_{e}\over{h}}\right)^{2}\left({I_{e}\over{I_o}}\right).
\end{equation}

Eqs.~\ref{eq:totalbulge} \& \ref{eq:totaldisk} should be multiplied by the
factor $(b/a)$ when considering projections on the plane of the sky.  One may
use Eq.~\ref{eq:bdrat} in a general sense, independent of projection, under
the assumption that the bulge and disk density distributions have similar
axes ratio (nearly true for late-type galaxies).
A weakness of $B/D$ ratios for systematic comparisons of galaxy light
profiles is its model dependence and the potential covariances between
some of the model parameters.  Consider Fig.~\ref{fig:sersicn} (top)
for the relative light fractions contributed by profiles of
different $n$ values, normalized to $n=1$.  The integrated bulge
light increases steadily as a function of $n$, for given values of $r_e$ and
$\mu_e$.  Thus, the adopted $n$ value in a bulge-to-disk decomposition has a
strong influence on the computed $B/D$ ratio.  Additionally, since larger $n$
profiles contribute light out to large $r$, the combination of a high $n$ and
a low $\mu_e$ (bright $r^{1/4}$ bulge) could take away light from the
outer disk and artificially boost the $B/D$ ratio.  A discussion on
non-parametric statistics, such as concentration indices
\citep{Kent85,Court96a,Graham01} which alleviate model dependences, is
presented in Paper~II\@.

\bigskip
We model the total galaxy luminosity profile as a sum of bulge + disk
components:
\begin{equation}
\label{eq:total}
I_{tot}(r)=I_{b}(r)+I_{d}(r).
\end{equation}
Profile smearing by atmospheric turbulence is accounted for in B/D
decompositions by convolving Eq.~\ref{eq:total} with a Gaussian PSF
of the form of Eq.~\ref{eq:seeing}.

Similar B/D analyses have also considered additional terms for a Gaussian
bar \citep{deJong96a}, a lens or ring \citep{PriAguVarMun01}, spiral arms,
and stellar disks with inner and/or outer truncations (Kormendy 1997;
Baggett, Baggett, \& Anderson 1998).  We restrict
our choice of fitting functions to a \sersic\ bulge and a non-truncated
exponential disk for a number of reasons. We find no prominent bars in our
sample and most our disk profiles are fairly linear (in magnitude space).
Azimuthal averaging for 1D profiles smoothes out spiral arm features (to a
different extent depending on whether the position angle was fixed or allowed
to vary in the profile extraction. Removal of spiral arm signatures
from the light profiles or images would require more time and effort than
is warranted by our analysis at this stage.)  We do not consider a sharp
inner disk truncation for a number of reasons: (i) unsharp masking techniques
reveal spiral structure from the inner disk into the galaxy center (Courteau
1992, 1996b; Elmegreen, Elmegreen, \& Eberwein 2001);
(ii) using HST images of inner disks, Carollo (1999) also finds evidence for
inner spiral structure and nuclear star clusters in the centers of early-
to intermediate-type spiral galaxies; (iii) {\it all} components
of the Galaxy have their peak surface brightnesses in the center
(\eg\ Wyse 1999).  Thus, at least some evidence suggests that spiral disks
reach in all the way to the center of late-type systems.  A lowering of
the disk central surface density may occur as stars get heated up into
a bulge by the action of a bar-like instability.  An exponential profile
with a {\it core} may thus be a reasonable description of Type-II SB
profiles.  We do not consider this approach here, but point out that
resolved B/D kinematics of nearby galaxies would provide a clear
indication whether stellar populations have been strongly depleted
and/or systematically scattered vertically into a bulge.  Some of our
galaxies show outer disk truncation (\eg\ Fig.~\ref{fig:trunc_dec}),
but see \S~\ref{subsec:rmax}.

The best-fit parameters of the (data$-$model) comparison are those which
minimize the reduced chi-square merit function, described in intensity
units as
\begin{equation}
\label{eq:chi_lum}
\chi^2_{\nu}={1\over{\rm{N-M}}}\,\,{\sum_{i=1}^{\rm{N}}}
\,\left[{I_{gal}(r_i)- I_{s}(r_i;h,I_{0},r_e,I_{e},n)\over \sigma_i }
\right]^2
\end{equation}
where N is the number of data points used, M is the number free parameters
(\ie\ N~-~M = $\nu \equiv$ Degrees of Freedom), and $\sigma_i$ is the
statistical intensity error at each pixel (2D) or surface brightness
level (1D).  From here on the $\nu$ subscript will be omitted and the
$\chi^2$ variable refers to a $\chi^2$ per degree of freedom (unless
otherwise specified).

The global $\chi^2$ of intensities is clearly dominated by the contribution
from the disk, virtually irrespective of the fitted bulge.  This effect
would be accentuated in galaxies with prominent features, such as spiral
arms, rings, or lenses, which are not accounted for in our pure
exponential disk models.  Cases are found where B/D decompositions with
significantly different bulge exponent $n$ values for a given profile have
nearly the same global $\chi^2$ value (see Figs.~\ref{fig:chi_n2} \&
\ref{fig:bulge_fits}).
Thus, in order to refine our parameter search for the best-fit bulge and
disk model, we compute a separate, {\it inner}, $\chi^2$ statistic out
to twice the radius
where the bulge and disk contribute equally to the total luminosity of
the galaxy ($r_{b=d} \equiv 2r(I_b=I_d)$).  We label this statistic as
\chisqrin\ (see Graham 2001 for a similar formulation).  For cases where
the bulges are so small that they never truly dominate the light profile
(\ie\ $r_{b=d}$ is undefined), we compute the \chisqrin\ out to the radius
at which $\nu =1$.

\subsection{Reliability of the Decomposition Results}\label{sec:reliability}

This section describes extensive testing of our bulge-to-disk decomposition
programs.
Artificial SB profiles and images were
created with a wide range of bulge profile shapes and exponential disks
including realistic noise and seeing effects.  Real galaxies are clearly
more complicated than the sum of two idealized mathematical functions,
but these tests provide a reasonable base for a global understanding of
the reliability and limitations of B/D decomposition algorithms.
The mock catalog of SB profiles and images will be used
to address the following questions:

\begin{itemize}
\item   How reliable and meaningful are the bulge-to-disk decompositions
        and fitted parameters?
\item   How crucial are initial estimates?  Are model fits always converging
        to the lowest $\chi^2$ minimum?
\item   How do seeing effects and sky subtraction errors affect the
        decompositions, and can they be properly accounted for?
\item   Are the small bulges in late-type disk galaxies sufficiently
        resolved to permit a reliable solution of the \sersic\ $n$
        parameter as a free parameter?
\end{itemize}

The literature abounds with investigations of profile fitting algorithms
based on artificial data, such as Schombert \& Bothun (1987; hereafter SB87)
who performed
double-blind experiments where one of the authors created mock
luminosity profiles and the other independently fitted the data.
The SB profiles combined a de~Vaucouleurs bulge and an
exponential disk.
Photon noise, at a level matching typical blue CCD performances, and a
systematic 0.5--3.0\% error of the sky background were
added to the profiles.
SB87 found that the simultaneous fitting of disk and
bulge using standard NLLS techniques could reproduce the input
parameters to within 10--20\% in cases where
galaxy profiles can be decomposed perfectly as the sum of a bulge
and disk (which fails for Type-II profiles.)
SB87 claim that a sky estimate uncertainty of up to
3\% does not affect their derived parameters,
but we find that sky errors as small as 1\% can have a significant effect
on the shape of the outer disk profile and the derived bulge and disk
parameters (see \S~\ref{subsubsec:sky} below).  SB87 did not consider other
fitting functions but recognized that bulges may not be adequately described
by the de~Vaucouleurs $r^{1/4}$ function.  \citet{AndSan94} later examined
the inadequacy of the $r^{1/4}$ functional form for the bulge (1D) profile,
and first established the double-exponential nature of late-type spirals.

2D B/D decomposition techniques, which exploit the full galaxy image, were
also developed and tested in similar fashion in the mid-nineties
\citep{ByuFre95,deJong96a}.  De~Jong performed extensive tests
with mock galaxies modeled as pure exponential bulges and disks,
exploring the effects of errors in the measured observables including the
seeing FWHM, sky background level, minor over major axis ratio, $b/a$, and
position angle, PA.  These observables were used as fixed input parameters to
the fitting routine, and de~Jong calibrated the effect
of measurement error on the determined parameters by decomposing the
artificial galaxies using erroneous values for each observable.
He concluded that: errors in $\mu_0$ are predominantly caused by
sky subtraction errors and can be as large as 0.1~\magarc; errors
in $h$ can reach 10\% and are dominated by sky background and
ellipticity measurement errors; bulge parameter errors, of order 20\%, are
controlled by the B/D size and brightness ratios.  Bright bulges
are most affected by seeing errors, and fainter bulges can also be
affected by sky background errors.

Our own investigation reaches similar conclusions and further extends
de~Jong's simulations.
We test for the robustness of the fitting procedure and accuracy
of the derived parameters with various values of the fit initial estimates,
seeing FWHM, sky value, and their errors, and -- unlike de~Jong (1996a) --
we model the bulge with a
generalized \sersic\ profile.  These simulations were initiated by
\citet{BroCou97} but are extended here in much greater detail, especially with
respect to the determination of the bulge shape parameter $n$.

\subsubsection{Simulated Profiles and Images}\label{subsec:sim_prof}

Our tests use a large set of artificial SB profiles and
images which span a wide range of the bulge, disk, and seeing parameters.
The mathematical forms of the bulge and disk components are those discussed in
\S~\ref{subsec:algorithm}.  Noise was added to the model profiles and images
from a Gaussian distribution with deviation representative of the
standard brightness errors of our luminosity profiles at a given
surface brightness level (see Courteau 1996a, Fig.~9; Paper II) .

One hundred SB profiles and fourty images with realistic noise were
created for each bulge, disk, and seeing combination.
Most of the simulated profiles and images had the same disk parameters,
\begin{center}
$\mu_0=20$~\magarc, \\
$h=12$\arcsec, \\
\end{center}
which are representative of a typical galaxy in our sample\footnote{Different
values of $h$ were also tested but the results do not change significantly.}.
For exponential ($n=1$) bulges, the structural parameters were selected
from
\begin{center}
$\mu_e$ = 16, 17,..., 22~\magarc, \\
$r_e$ = 0.1, 0.2,..., 3\farcs0 \\
\end{center}
corresponding to $B/D$ ratios ranging from 0 to 5 and $B/T$ from
0 to 0.8 (see Eq.~\ref{eq:bdrat}).

All profiles and images were convolved with a seeing disk of
\begin{center}
$\rm{FWHM}$ = 1.0, 1.5,..., 3\farcs0. \\
\end{center}
Our models span the full range of parameters typically found in
late-type bulges (\eg\ \citet{deJong96b}; \citet{Court96a};
\citet{BroCou97}) and the seeing values match the expected range
at Lowell Observatory and KPNO (see \S~\ref{sec:data}).

We also explored the following range of \sersic\ $n$ values
\begin{center}
$n$ = 0.2, 0.4,..., 4.0 \\
\end{center}
for the set of combinations with $r_e$ = 0.8, 1.5, and 2\farcs5,
$\mu_e$ = 18, 20, and 22~\magarc and for seeing FWHMs of 1.5, 2.0 and
2\farcs5.  Here the corresponding $B/D$ ratios range from about 0 to 1,
and $B/T$ from 0 to 0.5.  We also simulated the full range of $r_e$, FWHM,
and $\mu_e$ = 18, 20, 22, and 24~\magarc for $n=0.2$ and $n=4.0$ for a test
regarding the initial estimates (see \S~\ref{subsec:init_ests}).

A total of about 223,000 artificial SB profiles and 16,200
images were created and modeled.  The mock profiles were all sampled at
0\farcs5/pixel to match the data.

The 1D and 2D decompositions of the
artificial profiles and images were deemed satisfactory if they met the
following (rather liberal) criteria:

\begin{itemize}
\item solution found within 100 iterations
\item $0 < r_{e(fit)} < 500$\arcsec
\item $0 < \mu_{e(fit)} < 30$~\magarc
\item $0.05 < n_{(fit)} < 10$
\item $0 < h_{(fit)} < 100$\arcsec
\item $0 < \mu_{0(fit)} < 30$~\magarc
\end{itemize}

For each set of parameters, we require that at least two-thirds of the 100(40)
simulated profile (image) decompositions pass these criteria to be included
in the analysis.
The 2D tests were not developed as fully due to prohibitive computing times.
Thus our tests rely more heavily on the 1D technique, but we have confirmed
that the results from both techniques corroborate each other.

\subsubsection{Disk Initial Estimates}\label{subsubsec:mark_disk}

\begin{figure}
\plotone{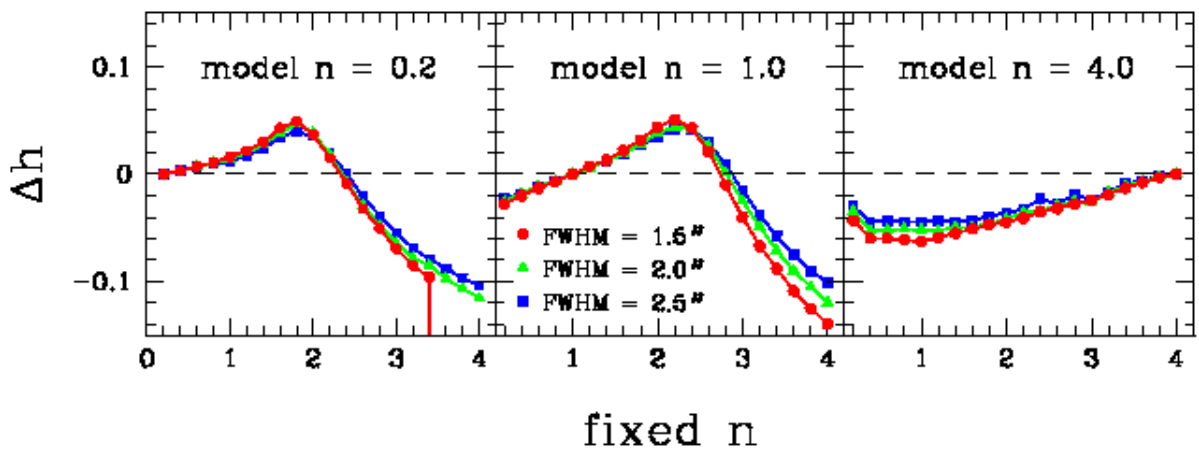}
   \caption{Effect of fitting an incorrect \sersic\ $n$ bulge on the disk
   scale length $h$.  Each panel plots the average relative fitted $h$ errors
   ($\Delta h \equiv ($\hfit(mean) - \hmodel)/\hmodel) with solid symbols and
   connected by solid lines as a function of the model $n$ for a bulge with
   $r_e=$2\farcs5 and $\mu_e=20$ \magarc. Red circles, green triangles, and 
   blue squares
   correspond to seeing values of 1.5, 2.0, and 2\farcs5 respectively.
   The three panels are for model $n$ values of 0.2, 1.0, and 4.0 from left
   to right. \label{fig:wrongn}}
\end{figure}

NLLS algorithms require initial estimates as input parameters, and we must
verify whether our final solutions are sensitive to our initial guesses.
Different initial estimates may yield different solutions
with comparable $\chi^2$ values especially if the topology of the $\chi^2$
distribution is non-trivial or shallow (\eg\ Schombert \& Bothun
1987\nocite{SchBot87}; de~Jong 1996a\nocite{deJong96a}).  Our simulations
confirm the robustness of our algorithms to a wide range of initial
\it{disk }\rm parameter estimates.  The results are slightly more
sensitive for large values of $n$, but in general the disk parameters were
perfectly recovered independent of the initial guesses.  One must still
caution that if the bulge is fit with the wrong $n$, the fitted disk
parameters will differ from their intrinsic values, even if the initial
estimates were good.  Fig.~\ref{fig:wrongn} shows the relative fit error for
$h$, $\Delta h$, where
\begin{equation}
\label{eq:deltah}
\Delta h \equiv {{h_{\small{\rm{fit}}}({\rm{mean}}) -
       h_{\small{\rm{model}}}}\over
      {h_{\small{\rm{model}}}}}
\end{equation}
versus the fitted $n$ value (held as a fixed parameter in the decomposition)
for bulges with $n$ = 0.2, 1.0, and 4.0.  (\ie\ $\Delta h=0$ indicates a
perfect recovery of the model parameter $h$).

In these decompositions the correct initial values were given for the bulge
and disk parameters and the seeing FWHM was assumed to be known.  Different
seeing values are represented by three point types: circles,
triangles, and squares for seeing values of 1.5, 2.0, and
2\farcs5 respectively.  We see that the fitted disk parameters can have
errors as large as 10\% when the bulge is modeled with an incorrect shape
parameter.  The reason for this is obvious if one considers the different
shapes for the different values of $n$ shown above in Fig.~\ref{fig:sersicn},
where the different shapes contribute differently to the outer profile.
The bulge $r_e$ is even more sensitive to the fitted $n$ (relative errors
in excess of 300\% for the worst cases, the figure is not shown).

\subsubsection{Bulge Initial Estimates}\label{subsubsec:bulge_init}

We now investigate the importance of the bulge initial
estimates.  Given that the fit baseline for late-type bulges is much smaller
than that of the disk and the apparent size of these bulges is comparable to
the seeing disk, the bulge model is likely to be much more sensitive to the
input parameters.
The sensitivity to the initial estimates of bulge parameters $r_e$ and $\mu_e$
was tested using offsets of $\pm$~25\% and $\pm$~50\% from the model, as well
as the correct model values as initial estimates.  This range of offsets
matches typical estimate excersions in our data modeling (see
\S~\ref{subsec:init_ests}).  The \sersic\ $n$ was
presumed known and the tests were performed for three different
values of $n = $0.2, 1, and 4.

The tests for the $n=1$ case demonstrated that bulge parameter initial
estimates are not important in both the 1D and 2D decompositions, as long as
$r_{e} \ga$ 0.3$\,*\,$FWHM.  Below this limit initial estimates
are more important; errors on $r_e$ can exceed 50\% and errors of up to
$\Delta\mu_{e}\pm0.3$ \magarc can occur.
In the $n=0.2$ case the parameter recovery is largely independent of the
seeing FWHM, as expected for profiles that are flat in the center.  Given
incorrect initial guesses, recovered parameters for profiles with
$r_e \ga$ 1\farcs0 can still be trusted, but for smaller bulges the algorithm
is trapped in a local minimum and the output value is nearly the
same as the input, \eg\ a $\pm 25(50)$\% input error yields a $\pm$25(50)\%
output error. For $n=4.0$ profiles, the parameter recovery is strongly
dependent on the seeing FWHM such that decomposition results for profiles with
$r_{e} \la$ 0.7$\,*\,$FWHM cannot be trusted.  Here again input and output errors
are approximately equal.  Moreover, even with correct initial estimates, the
model parameters are not perfectly recovered for profiles with
$r_{e} \la$ 1\farcs0.

We have interpolated these results for different values of $n$ and define
a parameter space for which our solutions are not affected by the choice of
initial estimates:

\[r_{e} \ga (0.3)^{1/n}\,*\,\mbox{FWHM} \]
and
\[r_{e} \ga \left\{ \begin{array}{ll}
            -0.75\,n+1.15 &   \mbox{ for $n \leq 1.0$}\\
            \,\,\,\,\,\,\,\,0.2\,n+0.2  &  \mbox{ for $n \geq 1.0$}
                        \end{array}
                \right. \]

The corresponding results for the 2D decomposition algorithm closely match
those from the 1D tests.

\subsubsection{Seeing Effects}\label{subsec:seeing}

\begin{figure}
\plotone{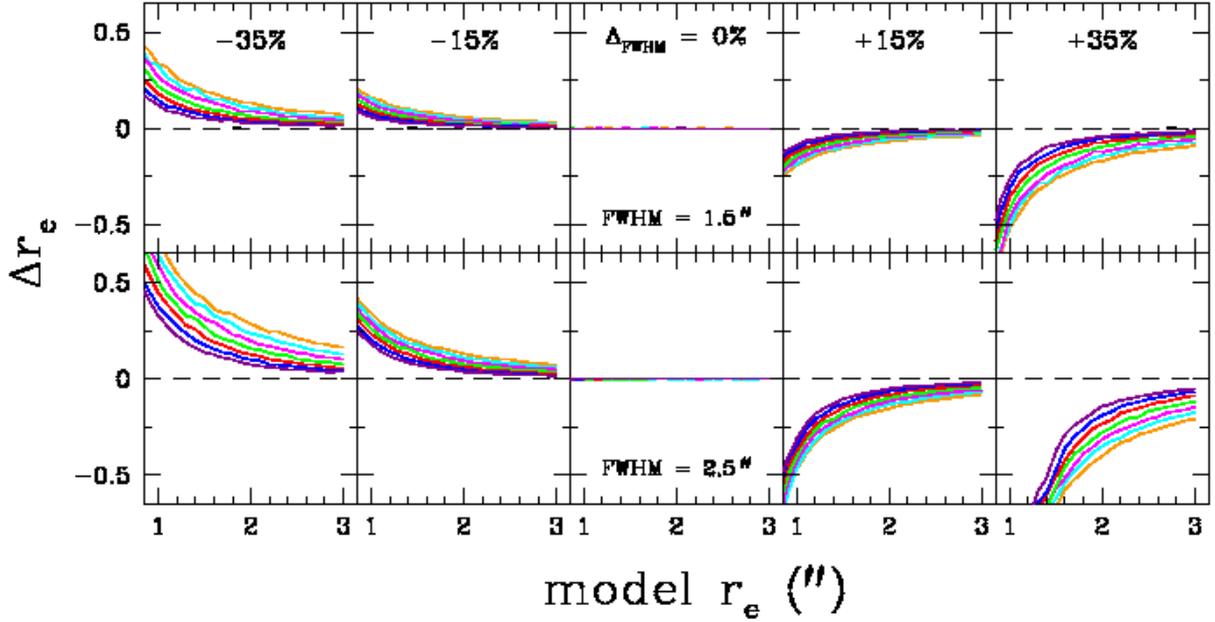}
   \caption{Effect on the fitted $r_e$ value of an incorrect seeing value in
    the 1D decomposition.  The column plots are based on different values
    for the fractional seeing error used in the fit, where
    $\Delta_{\hbox{\small{\rm{FWHM}}}} \equiv
    (\rm{FWHM}_{\hbox{\small{\rm{used}}}} -
     \rm{FWHM}_{\hbox{\small{\rm{model}}}})
     /\rm{FWHM}_{\hbox{\small{\rm{model}}}}$.
    Each row is for a different value of the model FWHM, 1\farcs5 (top) and
    2\farcs5 (bottom).  Each panel shows the average
    relative error on $r_e$,
    $\Delta r_e$ (Eq.~\ref{eq:deltare}), versus
    the model $r_e$.  The seven curves are for different values of $\mu_e$:
    16 (dark purple), 17 (blue), 18 (red), 19 (green), 20 (magenta), 21 
    (cyan), and 22 (orange) \magarc. \label{fig:seeing_test_re}}
\end{figure}

\begin{figure}
\plotone{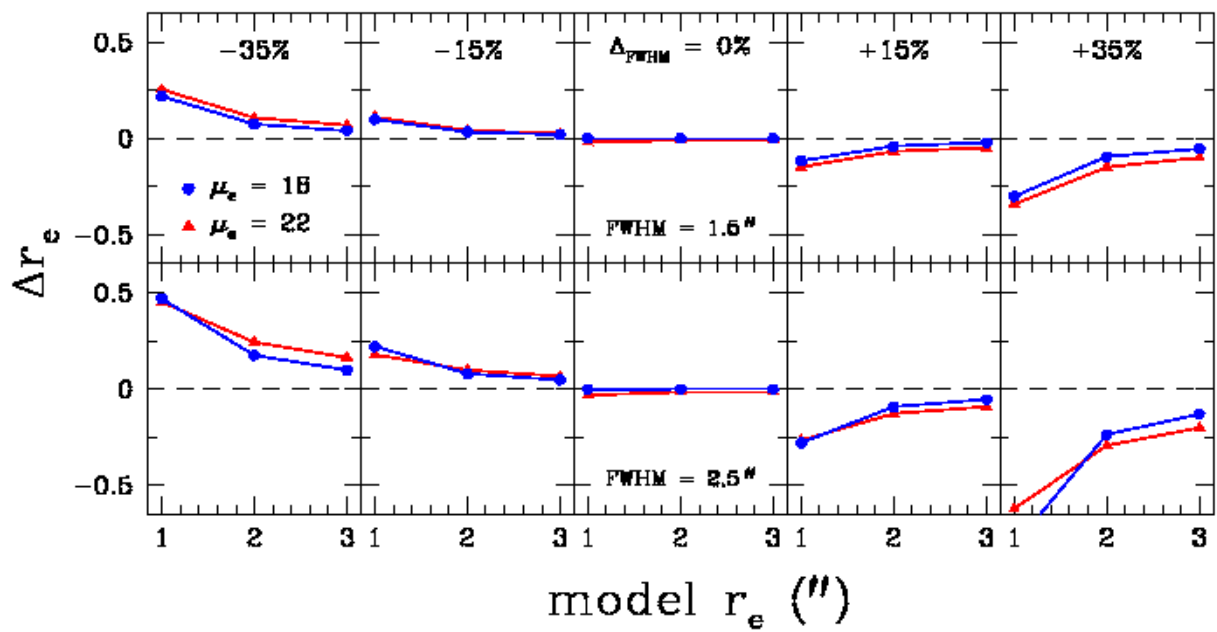}
   \caption{Effect on the fitted $r_e$ value of an incorrect seeing value in
    the 2D decomposition (compare with Fig.~\ref{fig:seeing_test_re}.)  The
    solid symbols connected by solid lines
    indicate the average (of the 40 image decompositions for each parameter
    and initial estimate combination) relative error on $r_e$, $\Delta r_e$
    (Eq.~\ref{eq:deltare}).  Blue circles and red triangles are for
    $\mu_e$ values of 18 and 22 \magarc respectively.  
    $\Delta_{\hbox{\small{\rm{FWHM}}}}$ is as defined in 
    Fig.~\ref{fig:seeing_test_re}
    \label{fig:2Dseeing_test_re}}
\end{figure}

The effect of an uncertainty in the seeing FWHM measurement on the model
parameters ($n$, $r_e$, $\mu_e$, $h$, and $\mu_0$) must be accounted for in
our simulations.  We fit each model using not only the nominal seeing value,
but we also varied the seeing FWHM by typical seeing measurement errors
(1 $\sigma$; $\sim 15-20$\% for optical, $\sim35$\% for infrared;
see Paper II).  Figs.~\ref{fig:seeing_test_re} \& \ref{fig:2Dseeing_test_re}
show the effect of an incorrect seeing estimate on the fitted $r_e$ for the
1D and 2D algorithms respectively.  Plotted are the relative fit error on
$r_e$, $\Delta r_e$, where
\begin{equation}
\label{eq:deltare}
\Delta r_e \equiv {{r_{\rm e, fit}({\rm mean}) -
       r_{\rm e, model}}\over{r_{\rm e, model}}} .
\end{equation}
The 1D and 2D simulations agree very well.
Figs.~\ref{fig:seeing_test_re} \& \ref{fig:2Dseeing_test_re} show that
when the correct seeing is used as input, the bulge and disk parameters
are recovered perfectly for the full range of bulge parameters and
for all values of the seeing FWHM tested.  However, even moderate seeing
uncertainties can severely affect bulge parameters depending on the size of
the bulge relative to the seeing FWHM.  If the seeing width is
under(over)-estimated $r_e$ is systematically over(under)-estimated,
worsening for smaller and fainter bulges and larger seeing values.  Similar
trends are seen for $\mu_e$.  Our tests show that the fit errors
can be significantly larger if the seeing FWHM is over-estimated than if it is
under-estimated.  As a rough rule of thumb, for $r_e \simeq \rm{FWHM}$ and a
seeing measurement uncertainty at the 35\% (15\%) level, the bulge $r_e$ can
be trusted to within 10--25\% (0--10\%), and $\mu_e$ to within $\pm$ 0.1--0.4
(0--0.2)~\magarc, the lower end of the range applying to the brightest bulges
and increasing towards the upper end for the fainter bulges.  For $r_e
\simeq \rm{FWHM} + 1$ the errors improve to within 0--15\% (0--10\%) for
$r_e$, and $\pm$ 0.0--0.2 (0.0--0.05)~\magarc for $\mu_e$.

There is no appreciable effect
due to seeing on the disk parameters (less than 1\%) except for the worst case
of a FWHM of 3\farcs0 and a 35\% seeing over-estimate.  In all other cases,
the disk parameters are virtually unaffected by seeing, as the size of the
disk is much larger than the seeing profile.  However, it is of paramount
importance to use accurate seeing
estimates and realistic seeing errors in order to sample the true range
of bulge parameters in B/D decompositions.

\subsubsection{Sky Uncertainty Effects}\label{subsubsec:sky}

We now test for the effects of an improper sky subtraction on the
decompositions.  The tremendous sensitivity of B/D decomposition
and scale length determinations to sky errors has been highlighted before
\citep{Court92,deJong96a}.  Here we aim to provide a firm quantitative
assessment
of such errors.  We re-model the same simulated profiles as in the
previous section but using sky values that are $\pm$1\% of the nominal sky
level (typical error in the optical passbands), and using a typical optical
sky brightness of 21~\magarc \footnote{Note that the percent sky error in the
H-band is much smaller than in the optical, but the sky at H is much brighter,
so the effect should be comparable.}.
Since bulge brightnesses are typically greater than the sky level, at
least at optical wavelengths, one might expect bulge parameters to be
somewhat insensitive to sky subtraction errors.  However, the outer disk
is very sensitive to sky subtraction errors and a modified disk
ultimately affects bulge structure due to their coupling.
Quantitatively, our tests show that if the sky is over- or under-estimated by
1\%, the error on the disk scale length, $\Delta h$, will be of order 5--15\%
and the disk CSB, $\Delta \mu_0$, will be $\pm$ 0.1--0.25~\magarc.  These
dispersions hold for the full range of bulge brightnesses except the two
faintest bulges which are one and two magnitudes fainter than the
sky; these had errors in excess of 50\%.  The
error ranges are controlled by the relative sizes of the bulge and
disk such that the disk parameter errors increase slightly from smaller to
larger bulges. This is simply because a larger bulge weakens the importance of
the disk in the central parts, thus giving more weight to the sky-sensitive
outer disk.

The errors on the bulge parameters are negligible for bulges with
$(\mu_{sky} - \mu_e) > 1$ \magarc for the entire range of $r_e$ and seeing
FWHMs, but increase up to $\Delta r_e \geq$ 15\% and $|\Delta \mu_e| \geq 0.1$
\magarc (increasing as the bulge gets smaller and as seeing conditions
degrade) for bulges with $(\mu_{sky} - \mu_e) < 1$ \magarc. In other words,
if the bulge effective surface brightness is less than one magnitude
greater than the sky brightness, the bulge parameters will be strongly
affected by sky subtraction errors. This effect is often neglected in
studies of bulge/disk structure.

The bulge and disk parameters are most affected for the case of an
under-subtracted sky.  This is largely due to a magnitude threshold of 26.5
\magarc in our decomposition algorithm.  The data are too noisy below this
value (Paper II) and we exclude them from the fits.
This threshold provides some protection against over-subtracted skies in the
measurement of the disk scale length.
Similar tests were performed with our 2D decomposition algorithm which
confirm, once again, the results above.

\subsubsection{\sersic\ n Tests}\label{sec:sersic}

\begin{figure}
\plotone{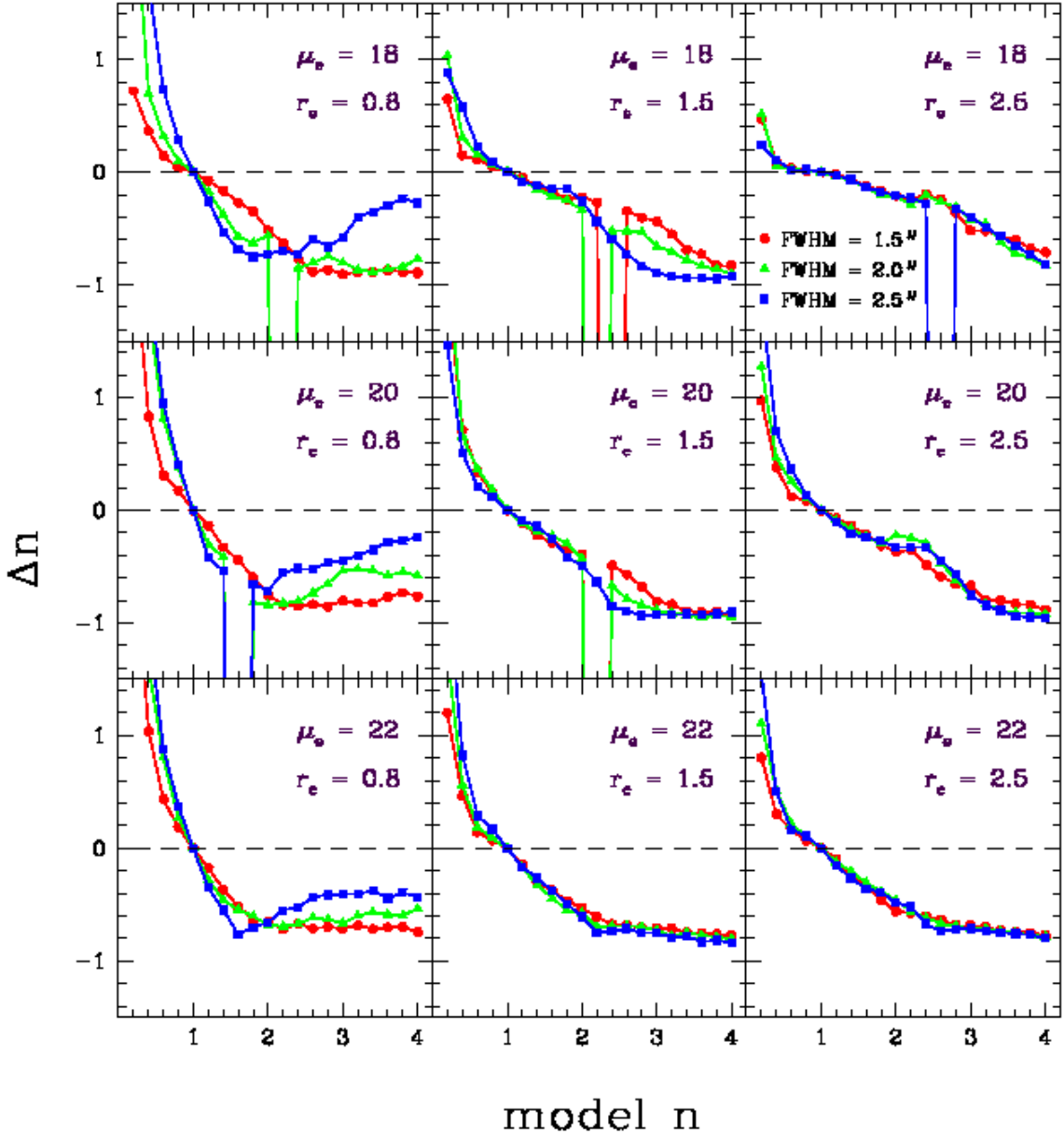}
    \caption{Difference between modeled and recovered values of $n$
    for a range of artificial profiles from
    $n=0.2-4$.  The \sersic\ exponent $n$ is a free fit parameter
    and the initial estimate is set to  $n=1$.  Each panel shows
    the average relative fitted $n$ errors ($\Delta n \equiv
    ($\nfit(mean) - \nmodel)/\nmodel) with solid symbols and connected by
    solid lines versus the model $n$ for the 9 combinations
    of $r_e=0.8$, 1.5, 2\farcs5, and $\mu_e = 18$, 20, 22 \magarc.
    Red circles, green triangles, and blue squares correspond to seeing
    values of 1.5, 2.0, and 2\farcs5 respectively.  The panels are ordered
    such that the $B/D$ ratio, for a given $n$ value, decreases from top to
    bottom and right to left panels. \label{fig:sersic_test1}}
\end{figure}

A number of recent studies have described the variation of bulge shapes
as a function of Hubble type \citep{AndPelBal95,MorGioHun98,KhoWadKem00,
MollHdt01}, going so far as suggesting that precise values of $n$
(\ie\ $\pm0.1$) could be determined \citep{Graham01}.  To our knowledge,
no study to date has tested the reliability of the recovery of
the \sersic\ $n$ parameter.
In order to test the sensitivity of the decomposition to the full
range of bulge profile shapes we use mock luminosity profiles with
values of the \sersic\ $n$ parameter ranging from $n=0.2$ to $n=4.0$.
The suite of profiles used all combinations of $r_e$ = 0.8, 1.5, and
2\farcs5, $\mu_e = 18$, 20, and 22~\magarc and seeing FWHMs of 1.5, 2.0,
and 2\farcs5.
The profile fits used initial estimates of $n = 0.4$, 1, 2, and
4, and correct initial estimates for $r_e$, $\mu_e$, and the disk parameters.
The seeing FWHM was fixed to the correct model value. The results for the
$n=1$ initial estimate are presented in Fig.~\ref{fig:sersic_test1}, where we
plot the average relative fit error on $n$ (for the 100 profiles with the same
simulated parameters) where
\begin{equation}
\label{eq:deltan}
\Delta n \equiv {{n_{\hbox{\small{\rm{fit}}}}({\rm{mean}}) -
       n_{\hbox{\small{\rm{model}}}}}\over
{n_{\hbox{\small{\rm{model}}}}}}
\end{equation}
versus the model $n$ (the dashed line at $\Delta n=0$ indicates a perfect
recovery of the model $n$ parameter.)  Each panel shows one particular
combination of $\mu_e$ and $r_e$, and the panels are arranged such that the
$B/D$ ratio, for a given value of $n$, decreases from top to bottom and right
to left.  The different seeing FWHM values are represented by three point
types: circles, triangles, and squares for seeing values of 1\farcs5,
2\farcs0, and 2\farcs5 respectively.

Fig.~\ref{fig:sersic_test1} reinforces that the bulges of even nearby
late-type spirals are small and not sampled at high enough spatial
resolution to yield a stable, robust solution for $n$ as a {\it floating}
parameter.  Given the
correct value of $n$ as an initial estimate (along with the correct initial
estimates for the other four parameters), the algorithm normally finds the
correct value of the model $n$, but any departure from the model value even by
a small amount, yields significantly different solutions for $n$, or the fit
may simply fail (as indicated by the vertical lines in the figures). For most
of the parameter combinations, an offset of $\sim$50\% in the initial estimate
of $n$ yields a $\sim$50\% error on its determined value.

Similar tests using the 2D algorithm show a slightly more robust recovery
of the model $n$ parameter based on incorrect initial estimates, but the
recovery efficiency is still poor and results based on a floating initial
estimate of $n$ are questionable.  We are here faced with an under-determined
optimization with too few independent data points for too many model
parameters (at least three for the bulge).  The strong covariances
between $n$, $\mu_e$, and $r_e$ ($\sigma_{n,\mu_e}, \sigma_{n,r_e}$)
prevent a unique determination of $n$ with this NLLS code.
We actually find the best-fit $n$ by grid search, holding $n$ as a fixed
parameter, solving for a range of values, and using the \chisqrin\ as
defined in \S~\ref{subsec:algorithm} to determine the best fit.  Further
simulations showed this technique to be fully reliable for the bulges
considered here.

It is difficult to estimate the error on $n$.  Either we use a grid search
and $n$ is fixed, or $n$ is kept as a floating parameter and varies widely
given wrong initial estimates.  Based on a test with floating $n$ but
correct initial estimates for bulge parameters, typical seeing and
sky errors modify $n$ by no more than 20\%.

Based on 2D B/D decompositions of simulated images and JHK images of 40
bright spirals, \citet{MollHdt01} estimate that recovery errors for all
the standard fit parameters ($I_d, h, I_e, r_e,$ and $n$) are less than
15\%, comparable to our findings.
They also tested for variable estimates of the sky level and seeing width,
though no clear description of their technique is presented.

\subsection{Summary of the Simulations}\label{sec:sim_summ}

The tests performed in \S~\ref{sec:reliability}, for idealized galaxies,
allow us to define a set of guidelines for the reliability and limitations
of our 1D/2D decompositions:

\begin{itemize}
\item   Initial estimates for bulge and disk parameters are unimportant
        provided that
\[r_{e} \ga (0.3)^{1/n}\,*\,\mbox{FWHM} \]

and

\[r_{e} \ga \left\{ \begin{array}{ll}
            -0.75\,n+1.15 &   \mbox{ for $n \leq 1.0$}\\
            \,\,\,\,\,\,\,\,0.2\,n+0.2  &  \mbox{ for $n \geq 1.0$}
                        \end{array}
                \right. \]

\item   Seeing errors must be accounted for in all bulge parameter studies.
        For $r_e \simeq \rm{FWHM}$ and a seeing measurement uncertainty at
        the 35\% (15\%) level, the bulge $r_e$ can be trusted to within
        10--25\% (0--10\%), and $\mu_e$ to within $\pm$ 0.1--0.4
        (0--0.2)~\magarc.
        For $r_e \simeq \rm{FWHM} + 1$ the errors improve
        to within 0--15\% (0--10\%) for $r_e$, and $\pm$ 0.0--0.2
        (0.0--0.05)~\magarc for $\mu_e$.  There is no appreciable
	effect due to seeing on the disk parameters (less than 1\%).
\item   Sky subtraction errors dominate disk parameter errors
        ($\sim $5--15\%) and are non-negligible (up to 25\%) for bulges
        whose effective surface brightnesses are less than one magnitude
        brighter than the sky brightness (\ie\ for $\{\mu_{sky}
        - \mu_e\} \la 1$).
\item   The sampling of late-type nearby bulges may not be high enough
        to constrain the \sersic\ $n$ exponent uniquely as a free parameter.
        Iterative model fitting schemes should be tested for this.
        Our approach uses a grid search.
\item   Typical seeing and sky errors modify $n$ by no more than 20\%.
\item   The 2D decomposition technique does not provide a significant
        improvement over the 1D method {\it for the recovery of axisymmetric
        structural parameters} to warrant the extra computational effort.
\end{itemize}

Armed with these basic guidelines we can now turn our attention to real
data decompositions using the 1D technique.

\section{Bulge-to-Disk Decompositions}\label{sec:decomps}

\subsection{Outline}
Our study of structural properties and the variation of galaxian
parameters as a function of wavelength uses the multi-band (BVRH)
data set of late-type spiral galaxies of Courteau, Holtzman, \& MacArthur
(\S~\ref{sec:data}; Paper II)\@.  Most galaxies have at least one set of
BVRH images,
and we use multiple observations for 54 galaxies to estimate systematic
errors.    Other B/D decomposition analyses have used larger samples
(\eg\ \citet{BagBagAnd98}) but lack the crucial multi-wavelength information.

We aim to develop a stable and versatile prescription to characterize
structural evolution of the bulges and disks of galaxies.
However, just as any morphological
description of galaxies (\eg\ Hubble types) depends on the wave band,
intrinsic structural parameters are also expected to vary with
wavelength due to stellar population and dust extinction effects.  Thus,
multi-wavelength information is required for any accurate description of
galaxian structural parameters.

Physical differences in the shape and size of bulges among galaxies are
also expected depending on how they were formed.  Formation by accretion
processes (\eg\ major/minor mergers) can account for steeply rising
de~Vaucouleurs light profiles in the central parts of galaxies (e.g.
van Albada 1982), while secular
evolution would yield exponential distributions, with or without a core,
of the central light. The formation
of small bulges is indeed largely attributed to secular processes
and redistribution of disk material (see \S~\ref{subsec:sec_ev}).
The present study is a natural extension of de~Jong's (1996a) structural
analysis of 86 face-on spirals with BVRIHK imaging, and Graham's (2001)
re-investigation of de~Jong's data.

De~Jong's 1D and 2D B/D decompositions
established significant parametric variations at different wavelengths.
Given the intrinsic limitations of the data modeling
(\ie\ over-determination of the parameter space), his B/D fits also
used a fixed \sersic\ $n$ parameter (see \S~\ref{sec:sersic}), but
limited to values of $n=1$, 2, and 4 bulges.  De~Jong's analysis, and that of
\citet{CourdeJBro96} who performed 1D profile decompositions for 290 r-band
luminosity profiles, supported the notion of exponential bulges and disks and
a tight correlation of B/D scale parameters in late-type spirals.
Evidence for this correlation was challenged by \citet{GraPri99} but later
validated by \citet{Graham01} who re-modeled de~Jong's thesis sample with
a 1D B/D decomposition technique\footnote{Graham's B/D analysis uses
an unconstrained (floating) S\'ersic shape parameter (see
\S~\ref{sec:limitations}).}.  His results support a range in the \sersic\
shape parameter from large ($n\simeq 2-3$) to small ($n\ga 0.5$) values for
early- to late-type spirals.
Aware of the inadequacy of basic B/D decompositions in fitting the bulge
shape parameter due to poor data resolution and strong covariances
with other bulge parameters
\citep[ see \S~\ref{sec:sersic}]{deJong96b,BroCou97},
we were compeled to revisit this issue with our own well-tested technique
and a more extensive data base.

Our approach involves B/D decompositions with fixed $n$ values that
sample the full parameter space of spiral bulges, from $n = 0.1,0.2,...,4.0$.
The fit solutions are filtered out on basis of relative \chisqr\ and
criteria based on our simulations (\S~\ref{sec:sim_summ}).  We are
concerned below with the derivation of robust B/D parameters for each
galaxy profile.  We compare our results with Graham (2001) and
others, and test for any B/D parameter correlations in \S~\ref{sec:results}.

The following is based exclusively on results from 1D B/D
decompositions.  The fact that we do not model non-axisymmetric
shapes (bars, rings, oval distortions) lessens the need for
more computationally intensive 2D B/D decompositions, as our simulations
showed no improvements using the 2D over the 1D decomposition method
for axisymmetric stucture.

\subsection{B/D Initial Estimates}\label{subsec:init_ests}

In order to determine the range of best fitted bulge and disk parameters,
we need to assist the minimization program in finding the lowest possible
(data$-$model) \chisqr.  From analysis of our mock images and profiles,
we have found that any reasonable initial estimates for the {\it disk}
parameters yields a robust solution.  We base our initial estimates for the
disk parameters $h$ and $\mu_0$ on the ``marking the disk'' technique, where
the linear portion of a luminosity
profile is ``marked'' and the selected range is fit using standard least
squares techniques to determine its slope.  Clearly, the resulting fits
are very sensitive to the adopted baseline.  We tested various choices for
the fit start and end points for our galaxy profiles including:
full profile fit, starting points of 0.2$r_{max}$ and 0.4$r_{max}$ out to
$r_{max}$, and a fixed baseline shifted along the length of the
profile and tracking 8 different locations.  Additionally, we also tested
the ``moments method'' of \citet{Willick99}.
The discrepancies between the different fits are large; ($\sim$10\% on average
and up to $\sim$100\% for the worst cases), but we found that the
0.2$r_{max}$ to $r_{max}$ baseline yielded the most reliable fits (as judged
by eye).  The inner boundary is chosen to exclude the major contribution of a
putative bulge or
Type-II dip and $r_{max}$ is the radius at which the surface brightness
error has systematically reached values greater than 0.12~\magarc (beyond
which the data become too noisy to be trusted).  The fits using the
0.2$r_{max}$ to $r_{max}$ baseline provided fits that were more than adequate
as initial estimates for the disk parameters in the decompositions.

Flexibility in the choice of bulge initial parameters is, however, only
afforded outside a certain range of bulge sizes relative to
the seeing disk.  Moreover, observed galaxy profiles show significantly
more variety than the idealized profiles from which these conclusions were
drawn (\eg\ we did not model Type-II galaxies, or the presence of strong
spiral features).  Accordingly, we explore three different sets of initial
bulge parameter estimates to protect against local minima in the parameter
space.  Initial bulge effective parameters were determined from:

\begin{itemize}
 \item Subtraction of the disk fit (based on the ``marking the disk''
       technique) from the original profile leaving only the bulge light.
       $r_e$ is then computed non-parametrically from the data by
       summing up the light up to the radius which encloses half
       the total light of the bulge.  Thus $\mu_e = \mu(r_e)$.
 \item $r_e =0.15h$ and $\mu_e=\mu_{0}$, where $h$ and $\mu_{0}$ are
       determined from the ``marking the disk'' technique.
 \item $r_e = {({b_n}/{\log(e)})}*0.15h$ and
              $\mu_e = (b_n - b_{n=1}) + \mu_{0}$.
\end{itemize}

The second set of initial estimates was motivated by the bulge/disk
structural correlation found by \citet{CourdeJBro96}
We added the third set of initial estimates which attempt to scale
$r_e$ and $\mu_e$ more appropriately to the different values of $n$.
No specific set of initial estimates worked better for all cases,
though the $3^{rd}$ method may be the least attractive.  It failed to provide
reliable solutions (\ie\ the fit failed or the $\chi^2$ values were large) in
most cases, but in a few cases it also yielded the only viable solution.

\subsubsection{Seeing and Sky Treatment}\label{subsubsec:singskytreat}

``Bulges'' of late-type spirals are small and their luminosity profiles can be
severely affected by atmospheric blur.  In principle, if the blurring from the
atmosphere (seeing) can be measured accurately, it can also be corrected
by Fourier deconvolution.  In practice, however, deconvolution
amplifies noise, and the seeing FWHM is subject to measurement
errors.  In
\S~\ref{sec:simulations} we used extensive simulations with a wide range of
input parameters and various values of $n$ to derive a space of recoverable
parameters under specific seeing conditions, accounting for the typical
measurement errors of our data.  Seeing is accounted for by convolving the
model profiles with a Gaussian PSF (Eq.~\ref{eq:seeing}) whose dispersion
is measured from field stars.  In order to account for seeing
measurement errors, each profile is modeled with three different values of the
seeing FWHM: the nominal measured value and $\pm$15\% of that value.  A mean
seeing uncertainty of $\pm$15\% was used rather than the individual errors per
measurement as these fluctuate greatly due, in large part, to the different
number of stars in each measurement.

Sky subtraction errors, of order $\sim 0.5$--1.0\% in the optical and
$\sim 0.02$\% in the H-band, were also examined carefully
(\S~\ref{subsubsec:sky}).  The sky brightness measurement error is
accounted for in B/D decompositions by using three different sky levels:
the nominal measured value and $\pm$0.5\% (optical) or $\pm$0.01\% (H-band),
of that value.

Each profile is thus reduced 3 times for each different combination of $r_e$
and $\mu_e$ initial estimates, times 3 seeing FWHM values,
times 3 sky values, and times 40 different fixed values of $n$, for a total
of 1080 decompositions per profile.

\subsection{Data Filtering}\label{subsec:filtering}

The 1080 decompositions for each profile are first vetted on the basis of
structural criteria determined from our simulations
(\S~\ref{sec:reliability}).  A decomposition is deemed acceptable if it meets
the following criteria:

\begin{itemize}
\item  $ r_{e} \ga (0.3)^{1/n}\,*\,\mbox{FWHM}\,\,\,\, $
 and
       $\,\,\,\, r_{e} \ga \left\{ \begin{array}{ll}
            -0.75\,n+1.15 &   \mbox{ for $n \leq 1.0$}\\
            \,\,\,\,\,\,\,\,0.2\,n+0.2  &  \mbox{ for $n > 1.0$}
                        \end{array}
                \right. $
\item   $B/D < 5$
\item   $h < 15$ kpc ; \,\,$r_{e} < 50$ kpc
\item   $r_e/h < 1$
\end{itemize}

The first constraint is derived from our simulations and
effectively eliminates small bulges whose sizes
are comparable to, or smaller than, the seeing disk.  The remaining
constraints are based on physical considerations and help eliminate
solutions with small \chisqr\ values but unrealistic parameters for
late-type galaxies.  Note, however, that these physical constraints
are rather generous and do not contribute any subjective bias.

The successful decompositions are then ranked on the basis of two indicators:
(a) a {\it global} \chisqr, \chisqrgl, computed for the full SB
profile from $r=0$ to $r_{max}$; and (b) an {\it inner} \chisqr, \chisqrin,
which includes only the central regions of the galaxy from $r=0$ to twice
the radius, $r_{b=d}$, where the intensities of the fitted bulge and disk are
equal (see \S~\ref{sec:method}).  \chisqrin\ was adopted to increase the
sensitivity of the goodness-of-fit indicator to the
bulge area\footnote{Note that our algorithm minimizes the \chisqrgl\ only.
The \chisqrin\ is calculated and used as a discriminator only after the
algorithm has converged.}.  The radius $r_{b=d}$ is clearly a function of the
bulge shape and may change from small to large $n$ (see \eg\
Fig.~\ref{fig:sersicn}).  Thus we use a \chisqr\ per degree of freedom to
remove any dependence of the normal \chisqr\ to a changing $r_e$.  Because of
the presence of spiral arms and other
non-axisymmetric features which we do not attempt to model, the reduced
\chisqr\ is always, in principle, greater than unity.  However, some of our
solutions may have \chisqr\ values less than unity indicative of an
over-determined system (correlated parameters), or over-estimated errors.

We first rank the solutions according to their \chisqrgl\ and preserve only
the better half.
The reduced set is then ranked according to \chisqrin\ values
and the bottom half of the distribution is discarded.  This process is
iterated at least twice, or until we reach 50 or fewer solutions.  Solutions
with \chisqrgl\ greater than 50 in this final subset are discarded.

Ideally, the minima for the distributions of \chisqrgl\ and \chisqrin\
values should agree to a common value of $n$, but differences may exist.
We search the final $\leq50$ solutions for a common solution,
starting at the minima of each  \chisqr\ distribution.  If the $n$ values
corresponding to the two \chisqr\ minima do not agree, the $n$ values
for the next smallest \chisqr\ values are compared (with the lower \chisqr\
values and with each other), and this process is iterated up to three times
until a match is found.  If this process did not converge, \ie\
there is no true minimum in the ((\chisqrgl,~\chisqrin) - $n$ space), a
final solution is chosen corresponding to the minimum value of
(\chisqrgl/min(\chisqrgl) + \chisqrin/min(\chisqrin)).

\begin{figure}
\plotone{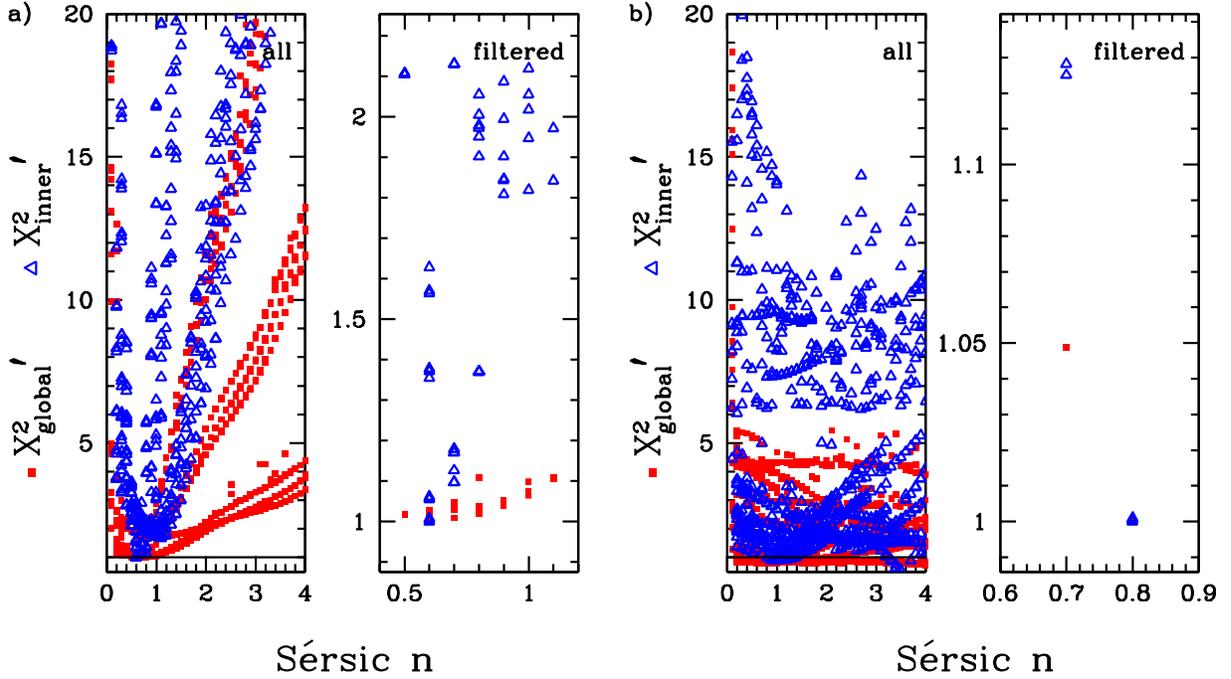}
    \caption{Examples of $\chi^{2}_{inner}\prime$ (open blue triangles)
    and $\chi^{2}_{global}\prime$ (filled red squares) versus \sersic\ $n$
    distributions for the 1080 decompositions of two different V-band
    observations of the same galaxy (UGC 929). In the two sets of plots
   (\,a) and b)\,),
    the left panel displays all 1080 points and the right panel shows
    only the ($\leq50$) points remaining after iterative filtering.
    Set a) shows a reasonably well-behaved solution
    favoring $n=0.6$ while set b) shows a rather
    noisy solution favoring $n=0.8$. \label{fig:chi_n1}}
\end{figure}

\begin{figure}
\plotone{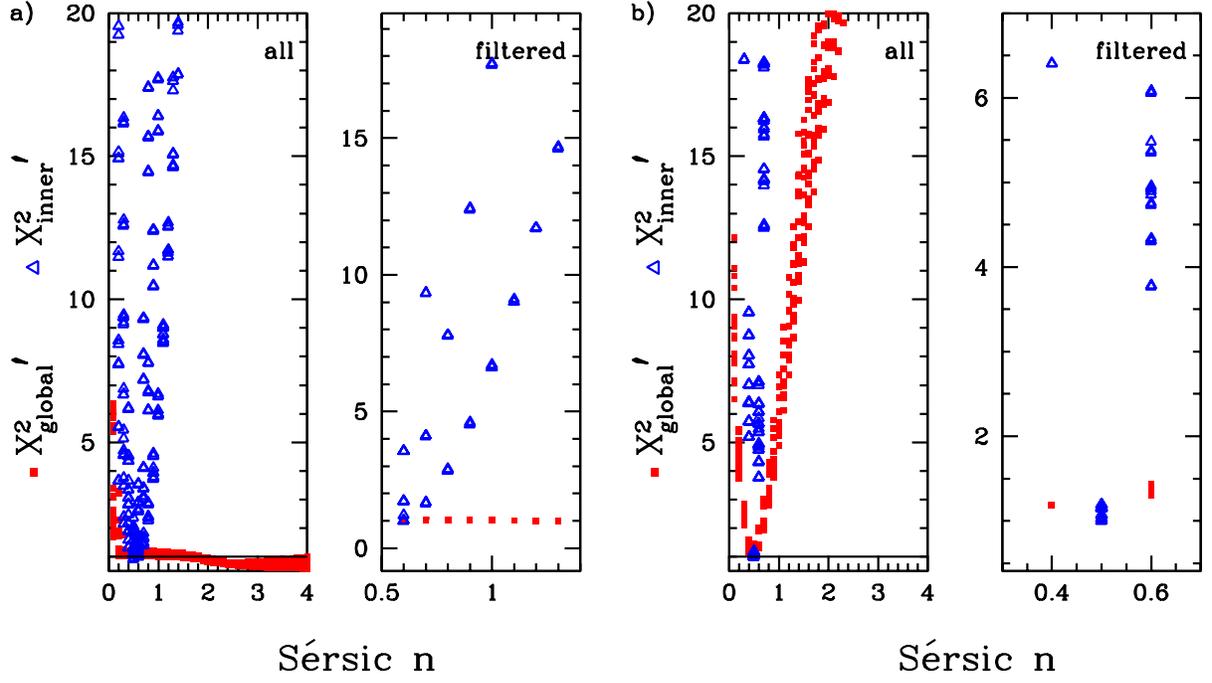}
    \caption{Examples of $\chi^{2}_{inner}\prime$
    and $\chi^{2}_{global}\prime$  distributions for a solution with a
    well-behaved $\chi^{2}_{inner}\prime$, but a flat
    $\chi^{2}_{global}\prime$ distribution (UGC 784 B-band), plot a), and
    for a very well-behaved solution in both $\chi^2$ distributions
    (UGC 929 B-band), plot b). \label{fig:chi_n2}}
\end{figure}

\begin{figure}
\plotone{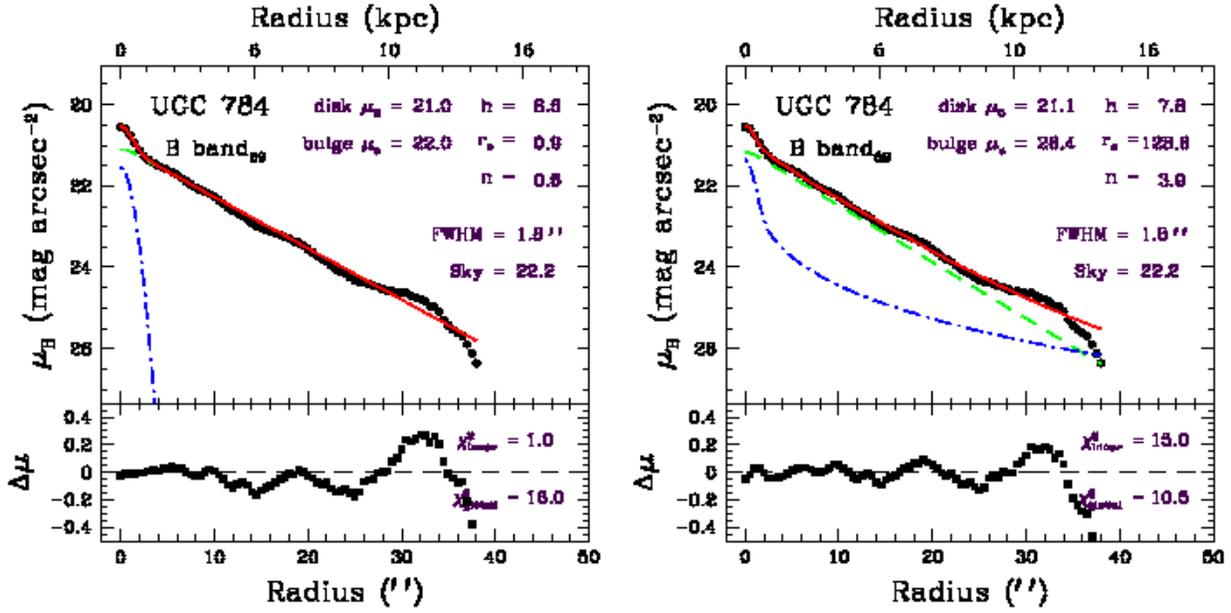}
   \caption{Comparison of different bulge fits for the same
            profile (UGC 784 B-band).  The plot on the right has a bulge fit
            (dashed-dotted blue line) which is likely unphysical.  Its
            \chisqrgl, however, is lower than that
            of the decomposition on the left plot, whose bulge fit looks more
            realistic.  Without adopting the \chisqrin\ statistic, the plot on
            the right is favored.  Using the \chisqrin\ in addition to the
            \chisqrgl\ as a discriminator, the plot on the left is favored.
            (See left plot of Fig.~\ref{fig:chi_n2} for the corresponding
            \chisqr\ vs. $n$ distributions.) Symbols, colors and line-types 
           are as defined in Fig.~\ref{fig:typeI_dec}. \label{fig:bulge_fits}}
\end{figure}

Figs.~\ref{fig:chi_n1} \& \ref{fig:chi_n2} show examples of the
distributions of (\chisqrgl,~\chisqrin) versus $n$ where
$\chi^{2}_{global}\,\prime
\equiv{\chi^{2}_{gl}}$/min$(\chi^{2}_{gl,filt})$ and
$\chi^{2}_{inner}\,\prime \equiv {\chi^{2}_{in}}$/min$(\chi^{2}_{in,filt})$,
where min$(\chi^{2}_{filt})$ is the minimum \chisqr\ value from the set of
($\leq50$) filtered solutions.  Note that these minima do not necessarily
correspond to the lowest value of the respective distributions from all 1080
solutions, as the initial absolute minima may have been filtered out
(\ie\ a poor combination of \chisqrgl,~\chisqrin\ for a given solution).
Thus, the normalized $\chi^2$'s may be less than one (as is easily seen in the
leftmost plot of Fig.~\ref{fig:chi_n2}).
In these plots, the left panels show the \chisqr\
distributions for all 1080 decompositions, while the right panels display
only the $\leq50$ solutions remaining after the iterative filtering scheme
described above.

Fig.~\ref{fig:chi_n1} highlights the sensitivity of our technique for two
V-band observations of UGC~929 taken under different seeing/sky conditions.
The left figures show a fairly well-behaved solution favoring $n=0.6$ and the
figures on the right plot show a rather messy solution favoring $n=0.8$.  The
seeing conditions were worse and the sky was much brighter for the observation
shown on the right which could explain the noisy distributions of both the
\chisqrgl\ and \chisqrin.

Fig.~\ref{fig:chi_n2} shows two different behaviors of \chisqrgl\ for profiles
with very well-behaved \chisqrin.  The plot on the left for our UGC 784 B-band
profile illustrates the need for an additional, more discriminating statistic
for the bulge region.  Decompositions based solely on the \chisqrgl\
goodness-of-fit indicator may result in fits, like the one shown on the right
side of Fig.~\ref{fig:bulge_fits} (dashed-dotted blue line).  However,
Fig.~\ref{fig:chi_n2}~a) clearly shows that the fit with $n=0.6$ is
a far superior match to the bulge shape, as indicated by the \chisqrin\
behaviour.

The final step of our filtering procedure entails a visual inspection of the
final decompositions.  The criteria for user examination include
information from multiple exposures and multi-band reductions for a given
galaxy. Profiles and/or solutions with the following pathologies were
eliminated from the final sample:

\begin{itemize}
\item  disk profiles that are too short for proper fitting
\item  no obvious, extended, underlying exponential structure for the disk
       (occurs predominantly in Type-II profiles)
\item  unphysically large fitted bulge
\item  unrealistic disk fit for Type-II profiles.  The fit is tipped
       below the true disk to account for the Type-II dip near the
       bulge-disk transition
       region leading to scale lengths that are biased high (\eg\
       see Fig.~\ref{fig:typeII_dec} for UGC 12527 for examples of
       ``bad'' fits which were eliminated from the final sample)
\item  large deviations between solutions for multiple
       observations of a given galaxy.
\end{itemize}

Not surprisingly, most of the eliminated profiles are Type-II systems.
We caution that even the Type-II profile decompositions that survived the
full sorting process may not provide the ideal description of their complex
surface brightness distributions.  Cleary these Type-II profiles
cannot be properly modeled with just a \sersic\ bulge and exponential disk.
Out of 523 images/profiles, a total of 341 passed our acceptance criteria.

\subsubsection{Preferred Sky and Seeing}\label{subsec:skysingoff}

Histograms of the preferred seeing FWHM and sky offsets for all decompositions
in all four bands are shown in Fig.~\ref{fig:skysinghist}.
Typically, a lower sky brightness level is preferred by our algorithm.
In some cases, this could be explained by an over-estimated sky level,
but it may also be due in part to profiles with
truncated outer disks as in Fig.~\ref{fig:trunc_dec}.  Our program prefers
a slightly larger seeing FWHM than measured.  This could be the result of
an under-estimated FWHM, or perhaps differences between
the idealized Gaussian model and the real seeing PSF.
Solutions with variable sky/seeing estimates were retained in the
final solution set for assessment of parameter errors.

\subsubsection{Effect of $r_{max}$}\label{subsec:rmax}

Of significance to the fit results is the maximum radius used in the
decompositions.  We have used the full observed profile out to radii
where the surface brightness errors reached above 0.12 \magarc.
There is no absolute definition to the edge, $r_{max}$, of a disk
and a different selection could yield different results.  To test the
sensitivity of our parameter determinations to the chosen value of $r_{max}$,
we re-decomposed the profiles as described above, but with a fit baseline
extending only to $0.75 \times r_{max}$.  A comparison of the results from the
two techniques, prior to eyeball filtering, shows good agreement and we
chose to keep the larger baseline to avoid discarding good data.

\subsection{Decomposition Examples}

There is no room for a full display of our catalog of final
decompositions, but a few examples are shown in
Figs.~\ref{fig:typeI_dec}--\ref{fig:nobulge_dec}.  The full catalog
of decomposition plots is available upon request from the authors.

In these figures, the solid black circles are the data points,
the black dots show the sky error envelope (from the measured sky error),
the dashed and dashed-dotted lines show the disk and bulge
fits respectively, and the solid line is the total (bulge+disk) fit.
The fits are all seeing-convolved using the best selected seeing values.
The bottom panel shows the fit residuals where $\Delta\mu(r)$ represents the
(data$-$model).  Fig.~\ref{fig:typeI_dec} shows an example of the
quintessential Type-I profile at all wavelengths.
Fig.~\ref{fig:typeII_dec} shows a
Type-II/Transition galaxy whose Type-II signature significantly weakens from
the optical to the infrared.
Fig.~\ref{fig:trunc_dec} shows a Type-I profile with an outer truncated
disk.  Such decompositions will presumably favor an under-subtracted sky
in attempt to align the inner and outer parts of the disk.  Here is an
example where our procedure with an over/under-estimation of the sky
and an infinite exponential disk model may not be adequate since the
outer disk truncation appears real (as detected in all four bands).
We also show an example of a nearly bulgeless system in
Fig.~\ref{fig:nobulge_dec}.  Our sample is divided into
52 Type-I, 53 Type-II and 16 transition systems, of which
18 truncated and 7 bulgeless disks are identified\footnote{The fraction
of a given galaxy class should be interpreted with
care since our sample is not volume-limited.}.

\subsection{Distribution of the \sersic\ $n$ parameter}

Fig.~\ref{fig:hist_n_all3} shows histograms of the \sersic\ $n$ parameter
for all the final fits (left) and good fits only (right) after
user-examination as described above.  The distribution of $n$
has a definite range, implying that not all late-type bulges are best
described by an exponential profile, but the mean value is very close to one.
This result agrees with \citet{Graham01} and recent N-body simulations of
galaxy evolution (\S~\ref{subsec:sec_ev}).

\subsubsection{Floating \sersic\ $n$}\label{sec:limitations}

In \S~\ref{sec:sersic} we showed that resolution limitations prevented
stable fitting of the \sersic\ shape parameter $n$ as a free parameter.
To illustrate the effect a floating $n$ can have on fitted parameters we
re-decomposed all of our galaxy profiles leaving $n$ as a free parameter
(\eg\ akin to Graham 2001).
The results are shown in Fig.~\ref{fig:hist_n_floatn} for three different
initial guesses for $n$ (0.2, 1.0, and 4.0).

The histograms of the resulting distributions of $n$ reveal a strong bias
towards the adopted initial estimate.  All 3 distributions show a large
peak at $n=0.1$, indicative of poor bulge fits.  The histogram for the
$n=1.0$ initial estimate looks somewhat similar to our own constrained
solution (Fig.~\ref{fig:hist_n_all3}), but this is somewhat fortuitous given
the closely-exponential nature of spiral bulges.  Note also the non-Gaussian
tail in Fig.~\ref{fig:hist_n_all3} is not reproduced in
Fig.~\ref{fig:hist_n_floatn} for the $n=1$ initial estimate case.
Fig.~\ref{fig:floatn} shows a comparison of the $\chi^2$ values from the
floated versus fixed $n$ solutions.  (When no suitable fit was
found, all parameters were set to 0 as indicated by the points lying on
the axes; note the large number of fit failures in the floated $n$
case.)  Note the large discrepancies in \chisqrin\ and \chisqrgl\ between
the two methods.  Thus, while the final distributions for the $n=1$ initial
estimate and our constrained $n$ procedure look similar, significant
differences may exist between individual decompositions.

\subsection{Error of a Single Measurement}\label{sec:repeats}

An important feature of any decomposition technique is the stability of
the final results for repeat observations of a given system.  Our sample
has 50 profiles for which multiple (two to four) observations exist,
allowing for a direct measure of the reliability of our decompositions.
Table~\ref{tab:rms} gives the mean and mean standard deviation of the five
model parameters from repeat observations with,
\begin{equation}
\overline{\sigma_x} = {{\sum_{N}\left\{\left[{{\sum_{n}(x_n-\overline{x})^{2}}
\over{n-1}}\right]_{N}^{1/2}\right\}}\over{N}}
\end{equation}
where $x$ is the fit parameter, $n$ is the number of observations for a given
profile, and $N$ is the number of profiles with repeat observations.
The average errors from repeat
observations of Type-I profiles are $\pm\,14$\% for $n$,
$\pm\,0.2$ \magarc for $\mu_e$, $\pm\,13$\% for $r_e$, $\pm\,0.05$ \magarc
for $\mu_0$, and $\pm\,3$\% for $h$.  Clearly, determinations of disk
parameters are much more stable than for bulges.  Error terms quoted below
correspond to the 1-$\sigma$ deviation, unless otherwise noted.

\subsection{Comparison with Other Authors}\label{sec:compare}

The overlap between our sample and de~Jong's thesis catalog \citep{dJvdK94}
amounts to only 3 galaxies.  Direct comparison of our SB profiles
shows excellent zero-point and overall shape agreement (Paper~II); however,
our B/D decompositions differ somewhat, as shown in Table~\ref{tab:us_vs_dJ}
(note that de~Jong uses fixed $n=1$).
Also shown in that table are decomposition parameters for the same galaxies
by \citet{Graham01} (same data as de~Jong, but using a range of $n$).  We
find scale length differences at the 10\%
level with de Jong and Graham, consistent with, or slightly
better than, typical variations between different authors \citep{KnapvdK91}.
A comparable dispersion is measured between the scale lengths of Graham
and de~Jong based on 82 R-band profile decompositions.  Graham's scale
lengths are, on average, smaller for small galaxies and larger for
big galaxies (apparent size) than de~Jong's.  We find systematically larger
disk scale lengths than de~Jong (based on only 7 comparisons.) We verified
that sky under/over-estimates cannot account for any difference with de~Jong.
De Jong's algorithm gives more weight to the outer part of the disk,
possibly explaining the shorter disk scale lengths.
For profiles with outer truncated disks or Type-II decrements, greater
weight in the outer parts favors the outer disk curvature and thus steeper
disk fits.

Bulge parameters between us and Graham match
reasonably well for the first two galaxies but differ substantially
for UGC 3140.  It is however difficult to establish trends based on
just 3 comparisons.  We can, instead, broadly compare our respective
distributions of \sersic\ $n$ with morphological type.  This is done
in Fig.~\ref{fig:sersic_morph} for comparison with Fig.~10 of
\citet{Graham01}.  The general features are similar, but we find a
wider range of $n$ values for the later-types possibly due to
the larger number of Scd/Sd galaxies in our sample.
Another favorable comparison of bulge parameters with Graham is shown
in Fig.~\ref{fig:rb_rd} (see \S~\ref{subsec:bdrat}).

We also find an overlap of two galaxies, NGC 3512 and NGC 7782,
with the sample of Baggett \etal\ (1998).  As with de~Jong and Graham,
disk parameters agree within 10\%.
Bulge parameters from Baggett \etal\ are missing for NGC 3512, and those
listed for NGC 7782 (both V band) differ quite substantially from ours.
These authors find $\mu_e=10.88$ and $r_e=0.2$ for a de Vaucouleurs bulge
and we have $\mu_e=20.1$ and $r_e=3.2$ for n=1 (best fit) or
$\mu_e=25.2$ and $r_e=83.8$ for n=4 (very bad fit with high
reduced $\chi^2$).  Note that seeing estimates were comparable.
Surprisingly, their $\mu_e$ is nearly 10
magnitudes brighter than their $\mu_0=20.3$ (we also find $\mu_0=20.3$)!
We find this pathology in a number of their bulge decompositions (see
e.g.\@ their Fig.~2) where the models often overshoot the data at the
center.

We conclude this section by noting that disk scale lengths between
us and other authors differ at the 10\% level.  Our bulge parameters
are also qualitatively consistent with those of Graham.

\section{Discussion}\label{sec:results}

Simulations of galaxy profiles and images (\S~\ref{sec:simulations}) and
careful B/D decompositions (\S~\ref{sec:decomps}) have led to a final set
of structural parameters for late-type spiral galaxies
(Table~\ref{tab:decomps} in Appendix B).
These data can now be examined for intrinsic structural variations
and sensitivity to dust and stellar population effects.
The outline of this section is as follows:  First, we verify in
\S~\ref{subsec:incdep} that our solutions are not affected by projection
effects.  We then discuss in \S~\ref{subsec:bdresults} B/D parameter
variations both in the context of profile type differences and wavelength
dependence.
In light of existing limitations in our modeling
of Type-II profiles, our conclusions will be based mostly on
properties derived from Type-I profiles.  These will enable us to
examine the viability of secular evolution models for disk galaxies
(see \S~\ref{subsec:sec_ev}).

\subsection{Inclination Dependence}\label{subsec:incdep}

In order to test for projection effects, we plot the distributions of
$\mu_e$ and $r_e$, as well as disk $\mu_0$ and $h$ as a function of
ellipticity, $\varepsilon = 1 - b/a$, in Fig.~\ref{fig:inc_all}.
The surface brightnesses are only corrected for Galactic extinction
and cosmological dimming (as in \S~\ref{sec:data}); thus the $\mu_0$
and $\mu_e$ values should be considered as upper limits (\ie\ effective
brightnesses are too low). No trends with ellipticity are seen, including
the \sersic\ $n$ parameter and ratio of disk scale lengths (not plotted).
Furthermore, Types I, II and Transition are not confined to any particular
inclination range showing that the Type-II phenomenon is not an accentuated
feature due to line-of-sight extinction (\eg\ Type-II galaxies are not
preferentially inclined with the plane of the sky).

\subsection{Bulge/Disk Parameters}\label{subsec:bdresults}

Table~\ref{tab:means} shows the range of fitted
parameters at BVRH wavelengths for all galaxy profile types (Type-I,
II, and Transition).   The number of Type-II and
Transition systems included in this table (\eg\ only 4 decompositions
for Transition galaxies in the B-band) is drastically reduced from
our original distribution as many of them did not pass our validity
criteria (\S~\ref{subsec:filtering}).  Note that the parameters for
Transition profiles at H-band broadly match those of Type-I's at
that wavelength.

The \sersic\ shape parameter $n$ for Type-I
galaxies is near unity, within the errors, for all wavelengths.
Thus, we advocate that the natural, intrinsic
distribution of the \sersic\ $n$ parameter for late-type spirals has a
mean near 1.0 (with $\sigma_n \simeq 0.4$; see
Fig.~\ref{fig:hist_n_all3}).
By all accounts, bulges of late-type spirals are well-approximated,
on average, by a pure exponential (luminosity/mass) density distribution.

The distributions of \sersic\ $\mu_e$ and $r_e$ are broad, indicative
of the range of bulge types in our sample.  Effective radii are typically
less than 1 kpc.  Those of Type-II profiles are even smaller and
seemingly better determined than Type-I's but this is predominently
an artifact of our limited 2-component modeling.  Examination of
Type-II profile fits shows that the model disk is typically shallower
(than the ``true'' disk), as the fit accounts for the fainter
bulge/disk transition dip, and bulge effective radii are naturally
confined to a smaller range.

The distributions of disk scale lengths and their ratios show a clear
decreasing trend as a function of wavelength (as noted by de~Jong
(1996b)\nocite{deJong96b}). This statistically significant effect,
detected for all profile types, can be explained either by a high
concentration of older stars and/or dust in the central regions of the disk.
Absorption by dust alone can account for the scale length
ratios that we measure (see \eg\ \citet{Evans94}, Fig.~5).  Evans'
models do not consider scattering but for the nearly face-on galaxies
considered here, its effects are negligible (Byun \etal\ 1994;
de~Jong (1996c)).  The color gradient analysis of \citet{deJong96c}
using stellar population and dust extinction models suggests, however,
that dust and metallicity play a minor role but that age is
be the dominant factor.  Preliminary analysis of our photometric data
with the latest stellar evolutionary and dust models (MacArthur \etal,
in prep.; hereafter Paper III) suggests a combination of effects.
The interpretation of color gradients is non-trivial and may ultimately
require a full spectroscopic investigation to convincingly disentangle
the effects of age, dust, and metallicity.

\subsubsection{B/D Scale Ratios}\label{subsec:bdrat}

In Fig.~\ref{fig:rb_rd}, we plot $r_e$ vs. $h$ for our Type-I decompositions
(solid symbols) and those of \citet{Graham01} for de~Jong's BRK Type-I
SB profiles (open symbols). This figure provides the
basis for a renewed discussion of the suggestion by \citet{CourdeJBro96}
of structural coupling between the bulge and disk of late-type galaxies.
The large dispersions in the $r_e^\lambda$ and $h^\lambda$
(see Table~\ref{tab:means}) nearly cancel out to yield tighter $r_e/h$
correlations.  For Type-I profiles we find
$\langle r_e/h \rangle \simeq 0.22 \pm 0.09$ at all wavelengths,
corresponding to $\langle h_{\rm bulge}/h_{\rm disk} \rangle =
0.13 \pm 0.06$ for $n=1$.  This result is also borne out in
the H-band Transition profiles (see Table~\ref{tab:means}).
For comparison, \citet{CourdeJBro96}
found $\langle h_{\rm bulge}/h_{\rm disk} \rangle \sim 0.10 \pm 0.05$
(or $\langle r_e/h \rangle = 0.15 \pm 0.08$)\footnote{The
study of \citet{CourdeJBro96} combined the r-band decompositions
of \citet{BroCou97} and the K-band decompositions de Jong's
(1996a) thesis study but no distinctions were made between
Type-I and Type-II profiles.}.
The latter result is also in agreement with Graham (2001)
who finds $\langle r_e/h \rangle = 0.2$ (no quoted dispersion, but it is
somewhat larger than ours judging from Figs.~\ref{fig:rb_rd} and
~\ref{fig:reh_vs_morph}) for early and late-type spirals in the K-band.
This is consistent with a scenario
where bulges of late-type spiral galaxies are more deeply embedded in their
host disk, than earlier-type bulges.  In such an ``iceberg'' scenario
(e.g. Graham 2001), bulges and disks can preserve a nearly constant
$r_e/h$ but show a great range of $\mu_e$ for any given $r_e$.

In Fig.~\ref{fig:reh_vs_morph}, we show $r_e/h$ as a function of
morphological type from our (solid symbols) and Graham's (open symbols)
decompositions.  A mild trend with Hubble type is seen with
$\langle r_e/h \rangle = 0.20 - 0.013(T-5)$ ($1\sigma = 0.09$),
ranging from $\langle r_e/h \rangle \sim 0.20$ for late-type spirals to
$\langle r_e/h \rangle \sim 0.24$ for earlier types.  Comparison of
our and Graham's decomposition parameters in Table~\ref{tab:us_vs_dJ}
shows that large deviations may exist, thus only our data points
were included in the fit of  $\langle r_e/h \rangle $ vs $T$ above.
Data for different bands scatter evenly about the mean line.
More data at earlier and later
types would be needed to firm up this trend.  It is nonetheless
remarkable that early and late-type systems are described by very similar
scaling relations, thus suggesting comparable formation and/or evolution
scenarios.

\subsection{Test of Secular Evolution in Late-Type
            Spirals}\label{subsec:sec_ev}

This work has provided confirmation of two important structural signatures
of spiral galaxies which must be addressed by models of structure formation:
\begin{itemize}
\item {\it The underlying surface brightness distribution of late-type
      spirals has a range for the \sersic\ $n$ parameter from 0.1--2, but
      is best described, on average, by a double-exponential
      model of bulge and disk,} such as found in Type-I profile
      galaxies.

\item{\it Bulges and disks of late-type spirals are coupled,
      with $\langle r_e/h \rangle = 0.22 \pm 0.09$, or
      $\langle h_{\rm bulge}/h_{\rm disk} \rangle$ \ $= 0.13 \pm 0.06$,
      at all wavelengths.  A mild trend with Hubble type is also
      detected with a range $\langle r_e/h \rangle \sim$ 0.20--0.24,
      from late to early-type spirals.}
\end{itemize}

The first result describes the {\it large-scale} appearance of bulges.
Analyses of HST images have shown that a significant fraction of bulge
nuclei have power-law profiles ($r<500$ pc; \eg\ Phillips \etal\ 1996;
Balcells 2001) and host a central compact source (Carollo 1999).
The extent of these nuclear sources ($<$ 0\farcs3 for $cz<2500$ \kms) is
smaller than our images' pixel size and smoothed out by seeing.
We thus ignore their effects on the bulge light profile in this analysis,
but caution that our bulge parameters are to be considered upper limits if
a significant nuclear component is present.

A natural interpretation of the near constancy of B/D size ratios in
late-type spirals is that their bulges formed via secular evolution of
the disk. This scenario is possible if disks are bar-unstable,
which can be triggered by the global dynamical instability of a
rotationally supported disk or induced by interactions with a
satellite and if significant angular momentum transport is feasible
(\eg\ Martinet 1995; Combes 2000; see the collection of papers in
  Carollo, Ferguson \& Wyse 1999 for comprehensive reviews).
For bar-unstable disks, in particular to vertical deformations,
the inner disk material is heated up to 1--2 kpc above the plane
into a ``bulge'' via resonant scattering of the stellar orbits by the
bar-forming instability.  This in turn, catalyzes funneling of
disk material into the central regions and generates outward
transport of disk material in the outer parts.  Gas flows must also
be invoked to explain the higher spatial densities of bulges compared
to the inner disk.  Such a model is expected to produce correlated scale
lengths and colors between the disk and its central regions, as observed
(e.g. Terndrup \etal\ 1994; Peletier \& Balcells 1996; Courteau 1996b).
A ``bulge-like'' component with a nearly
exponential profile is expected from non-axisymmetric disturbances
that induce inward radial flow of disk material (Pfenniger \& Friedli
1991; Zhang \& Wyse 2000, and references therein).
The longer the disk-bar heating interaction, the greater the extent
of the disk exponential profile (Valenzuela \& Klypin 2002).
The evolving exponential attractor
is an empirical result well established in simulations, but it
lacks, at present, a theoretical explanation (Pfenniger 1999).

Although a bar can grow spontaneously (\lapprx 20 Myr) from
small scale fluctuations in the inner disk, an external finite
perturbation can catalyse its growth.  However, collisionless
mergers seem unsuited to growing the exponential bulges of
present-day late-type spirals, though they may contribute to the
increase in \sersic\ $n$ parameter seen toward earlier types in
proportion to the accreted satellite mass (Barnes 1988, Aguerri \etal\ 2001).
The spontaneous or triggered formation of bars also suggests that the
Hubble type of galaxies can change well after the formation of
the disk (Pfenniger 1999).  All of our bulges are smaller than
a disk scale length and could be created by purely bar-related
processes.  Instead, accretion of galaxy satellites is required
to make bigger bulges, either before or after formation of the host disk.

Secular evolution models of stellar and gaseous disks, especially through
cosmologically-motivated three-dimensional N-body simulations, have seen
significant developments in the last decade.
For example, the cold dark matter (CDM) hierarchical hydrodynamical
simulations by \citet{Saiz01} and \citet{ScaTissera02} show that secular
processes can occur naturally during the formation of spiral disks
and play an important role in the regulation of star formation and the
determination of the dynamical and structural properties of these systems.
On average, the simulated disk systems are shown to be characterized by
a double exponential profile which naturally emerges within the hierarchical
clustering scenario.  These results are based on a stellar formation
process implemented in such a way that it succeeds in forming compact
bulges that stabilize disk-like structure allowing the conservation of
an important fraction of their angular momentum during the violent phases
of their assembly.  Fig.~\ref{fig:tissera} shows the distribution of final
\sersic\ $n$ parameters for relaxed present-day late-type disks by
\citet{ScaTissera02}; their models reproduce our results
(Fig.~\ref{fig:hist_n_all3}) very nicely.
The double-exponential structure of bulge and disk may not always be
the final relaxed state of an object, but whenever $n \sim 1$, the
B/D scale ratio $\langle h_{\rm bulge}/h_{\rm disk} \rangle$ takes its
nominal value of 0.15.  These models do not have bulges with $n<0.7$,
possibly due to limited resolution and/or excessive angular momentum
transfer that supernova feedback could help prevent.

Simulations by Pfenniger (2002; private comm.\@) of self-gravitating disks
forming bars which may later dissolve into a bulge-like component also
show a nearly universal ratio $r_e/h$, in agreement with observed
values, which is related to the stellar dynamics of the barred system
(i.e.\@ relative position of the vertical to horizontal resonances).
The bar length is related to the initial rising part of the rotation
curve (yielding a scale), and the corotation of bars is proportional
to their length.  The corotation fixes the positions of the other
resonances, which in turn fix the maximum extension of bulges made
from resonant heating, as indeed the vertical resonances are strong
only within the bar.  This mechanism would thus set the upper limit
for the allowed range in $r_e/h$.

The $N$-body simulations of Aguerri \etal\ (2001), which consider the
growth of galactic bulges by mergers, also suggest that the final B/D
scale ratio $\langle r_e/h \rangle$ does not scale with the B/D
luminosity ratio.  These authors show that the disk
scale length $h$ can increase from 15\% (low mass retrograde satellite)
to 65\% (high mass direct satellite) while $\langle r_e/h \rangle$ would
decrease from 0.21 to 0.14, or 33\%, in the most extreme case.
One can thus infer $\langle r_e/h \rangle = 0.17 \pm 0.03$, independent
of B/D luminosity ratio, in good agreement
with our findings. Their simulations are however limited to a
small range of initial $r_e$ and a more complete investigation
with a broad range of $r_e$ and $h$ values is needed to establish
the fundamental nature of the B/D scale ratio.

\subsection{Type-II Profiles}\label{subsec:typeII}

The above scenarios for secular evolution naturally produce the
double-exponential character of the bulge and disk radial luminosity profiles
for late-type systems.  However,
over half our sample of 121 late-type spiral galaxies show strong
deviations from this simple two-component description.  Other authors
\citep{Kormendy77,BagBagAnd98} have considered inner disk truncation
(plus de~Vaucouleurs bulges) as an alternative to modeling Type-II
light profiles, with
\begin{equation}
I_{disk}(r) = I_\circ\exp\left\{-\left[r/r_\circ +
             (r_{hole}/r)^n\right]\right\}
\end{equation}
where $r_{hole}$ is the truncation radius and $n \sim 3$. As discussed
in \S~\ref{sec:method}, we do not consider this approach at the present,
but its potential merits should not be overlooked.

N-body simulations (\eg\ Norman, Sellwood, \& Hasan 1996;
Valenzuela \& Klypin 2002) reproduce Type-II surface density profiles
as a result of the redistribution of central stars into a ring by a bar-like
perturbation.  Approaching the centers where the component called bulge
and the component called exponential disk overlap,
one cannot tell, in these simulations, if a star or particle belongs
to which component.  Galaxy centers may recurrently move from a barred
to an unbarred phase and undergo continuing bulge building as
the bars dissolve\footnote{Only progressively larger bars in
the centers of exponential bulges would be allowed to form in a recurring
scenario due to the disrupting dynamical effect of a growing nucleus
(Rix 1998, as reported in Carollo 1999).} \citep{Norm96}.  Thus, the
paucity of barred galaxies in our sample does not preclude bar-induced
effects as a possible explanation for Type-II profiles
(e.g. Gadotti \& dos Anjos (2001)\nocite{GaddosAnj01}).
Pre-existing bars may simply have dissolved.  For example, out of 8
barred-classified galaxies in our sample, 6 have Type-II profiles
thus lending some credence to the bar-lens scenario.  On the other hand,
the most strongly barred galaxies in the Shellflow sample of $\sim 300$
bright late-type galaxies (Courteau et al.\@ 2000\nocite{Shellflow}) have
mostly Type-I profiles, indistinguishable in shape and global properties from
the profiles of unbarred Type-I galaxies (Courteau et al.\@, in prep.\@).
>From an inhomogeneous sample of 167 spiral galaxies, Baggett \etal\ (1996)
\nocite{BagBagAnd96} find only a weak tendency for barred galaxies to have a
higher occurence of Type-II profiles.  The link between Type-II profiles and
barred galaxies is thus unsecured at present.

Type-II profiles may also be explained by extinction effects in the disk.
Increased opacity towards the central disk can cause a depression in the
luminosity profile, especially at shorter wavelengths.  Realistic Type-II
profiles (in shape and colors) have been produced with
exponential distributions of stars and dust and variable layering
parameters \citep{Evans94}.
If dust extinction causes the inner disk profile dip, Transition galaxies
would just be a case of lesser dust content, whereas bona fide Type-II
systems remain optically thick, even at H-band.
Using far-infrared (FIR) to B-band flux ratios, and radiation transfer
models for the dust \citep{Gordon01}, we have tested for the origin
of Type-II signature as being due to extinction.  The FIR/B flux ratio should
be higher for the dustier systems.  Unfortunately, our measured total FIR/B
flux ratios are statistically identical (with large scatter) for Type-I,
Type-II, and Transition galaxies (Paper III), thus thwarting any clear
interpretation.  The IRAS 60 and 100 $\micron$ fluxes have too low
resolution and too large errors to separate the inner disk dust emission
from the whole galaxy.

If stellar population effects are relevant \citep{Prieto92}, age/metallicity
gradients should be detected at the bulge/disk transition in Transition
systems.  We will further investigate the dust and/or stellar population
origin of the Type-II dip in Paper III.

\section{Summary and Concluding Remarks}\label{sec:summary}

This study has focused on the development of rigorous B/D decomposition
techniques using a new, comprehensive, multi-band survey of late-type
spiral galaxies.  We examine three types of SB profiles, Freeman Type-I
and Type-II, and a third ``Transition'' class for galaxies whose profiles
change
from Type-II in the optical to Type-I in the infrared.  This distinction
is important since Type-II and Transition profiles cannot be adequately
modeled by a simple two-component model of the bulge and disk.  Thus,
our main results are based on Type-I profiles.

Based on extensive simulations, careful treatment of sky and seeing
measurement errors, and repeat observations we are confident that
systematic errors are $\la20$\% for the bulge components, including
the \sersic\ shape parameter, and $\la5$\% for disk components.

The main conclusions from our simulations and final profile decompositions
are as follows:

\begin{itemize}

\item Simulations to determine the range of acceptable solutions for
      any B/D decomposition program are crucial.  The reliability of bulge
      model parameters is limited by the relative size of the bulge and
      seeing disk, seeing errors, the intrinsic bulge shape, sky brightness
      and errors.  Disk parameters are fairly robust to systematic errors,
      with the exception of improper bulge shapes and sky errors which can
      have dramatic effects on both modeled disk and bulge components.
\item The \sersic\ bulge shape parameter for nearby late-type galaxies
      shows a range between $n=0.1-2$, but, on average, their underlying
      surface brightness distribution is best described by a
      double-exponential model of bulge and disk.
\item Disk scale lengths decrease at longer wavelengths, indicative of
      a higher concentration of older stars and/or dust in the central
      regions relative to the outer disk.
\item We confirm and reinforce the result of \citet{CourdeJBro96} of a
      structural coupling between the bulge and disk of late-type
      spirals.  We find $\langle r_e/h \rangle = 0.22 \pm 0.09$, or
      $\langle h_{\rm bulge}/h_{\rm disk} \rangle = 0.13 \pm 0.06$,
      independent of wavelength.   A mild trend with Hubble type is
      observed with $\langle r_e/h \rangle = 0.20 - 0.013(T-5)$
      ($1\sigma = 0.09$), ranging from $\langle r_e/h \rangle
      \sim 0.20$ for late-type spirals to $\langle r_e/h \rangle
      \sim 0.24$ for earlier types.
      These results are consistent with scenarios of bulge formation
      in which bulges of late-type spiral galaxies are more deeply
      embedded in their host disk than earlier-type bulges.  Under this
      ``iceberg'' scenario, bulges and disks can thus preserve a nearly
      constant $r_e/h$ but show a great range of $\mu_e$ for any given $r_e$.
      The observed scale ratio is consistent with numerical simulations
      of self-gravitating disks and probably related to the stellar
      dynamics of an actual or pre-existing barred system.

\item The inner brightness profile signatures of Type-II galaxies are
      likely explained by a combination of dust extinction and stellar
      population effects and perhaps linked to the occurence of a bar,
      but no decisive conclusion can be derived at present.
\end{itemize}


\newpage

\acknowledgments

We are grateful to Marc Balcells, Eric Bell, Roelof de Jong, and
Daniel Pfenniger for their comments on earlier versions of this manuscript.
Cecilia Scannapieco and Patricia Tissera are also thanked for sharing
their material (Fig.~\ref{fig:tissera}) in advance
of publication.  Alister Graham kindly provided tables of his profile
decompositions of de~Jong's SB profiles for comparison
with our and de~Jong's similar results.  We also wish to thank the
anonymous referee for suggestions and comments that helped improved
the presentation and content of the paper.
This research has made use of the NASA/IPAC extragalactic
database (NED) which is operated by by the Jet Propulsion Laboratory,
California Institute of Technology, under contract with the National
Aeronautics and Space Administration.  LM and SC acknowledge financial
support from the National Science and Engineering Council of Canada.

\appendix

\section{Functional form for the \sersic\ $b_n$
parameter}\label{appendixA}

Eq.~\ref{eq:bnexact} cannot be solved in explicit closed form for
$b_n$.  Many of the numerical and analytical solutions found in the literature
agree well for $n>1$ but differ significantly for smaller values of $n$.
Fig.~\ref{fig:bncomp} shows a comparison of the two most commonly used
approximations (short- and long-dashed curves) with the exact solution for
$b_n$, computed to a numerical precision of one part in $10^7$ for all
$n \leq 10$ (see also Fig.~2 in Graham 1999).

As we wish to test for spiral bulges with \sersic\ $n$'s as small as 0.1, we
have adopted a formalism that is valid for all $n$.  To maintain
computational simplicity, and ensure a suitably accurate solution we
found it practical to divide the curve into two segments.
For all $n>0.36$ we use the asymptotic expansion of \citet{CiottiBer99} up to
O$(n^{-5})$ (their Eq.~18),
\begin{equation}
b_n \sim 2n-{1\over{3}}+{4\over{405n}}+{46\over{25515n^{2}}}
+{131\over{1148175n^{3}}}-{2194697\over{30690717750n^{4}}}+O(n^{-5})
\end{equation}
which is good to better than one part in $\sim 10^4$ in that range.  However,
for $n\leq 0.36$ this solution diverges.
Due to the rapidly changing curvature in the gamma function, and thus $b_n$,
at small $n$, it would be necessary to use an unrealistic number of terms in
the asymptotic expansion to achieve the desired accuracy.  For
$n \leq 0.36$, we find the best fitting polynomial of the form
$b_n = \sum_{i=0}^{m}a_i*n^i$ where $m$ is the order of the polynomial and
the $a_i$ are the coefficients of the fit given by,
\begin{equation}
\begin{array}{lllll}
a_0=0.01945 & a_1=-0.8902 & a_2=10.95 & a_3=-19.67 & a_4=13.43.\\
\end{array}
\label{eq:ourbn}
\end{equation}
This fit is accurate to better than two parts in $10^{3}$.
The wiggles in the dotted curve in Fig.~\ref{fig:bncomp} result from
the polynomial nature of the fit and limited numerical precision where
the gamma function approaches infinity.

\bigskip

\section{Decomposition Results for the Type~I
   Profiles}\label{appendixB}
Table~\ref{tab:decomps} gives relevant photometric information and 1D B/D
decomposition results for the final set of Type I galaxy profiles.
Decomposition results for Type-II and Transition galaxies are available
from the authors upon request, with the caution that parameters
for these profile types should be interpreted with care.  The
entries are arranged as follows:

{\it Column (1)}: (UGC number) (observation number) (passband) for
each profile;

{\it Column (2)}: Ellipticity,  $\varepsilon\equiv(1-b/a)$.  The final
ellipticity (and position angle) estimates correspond to an average of
those values from the five contours surrounding the best isophotal fit
in the outer disk, as determined by eye. This estimate is clearly
sensitive to the presence of spiral arms.  The typical inclination error
is $\sim 3\deg$, independent of ellipticity;

\begin{figure}
\plotone{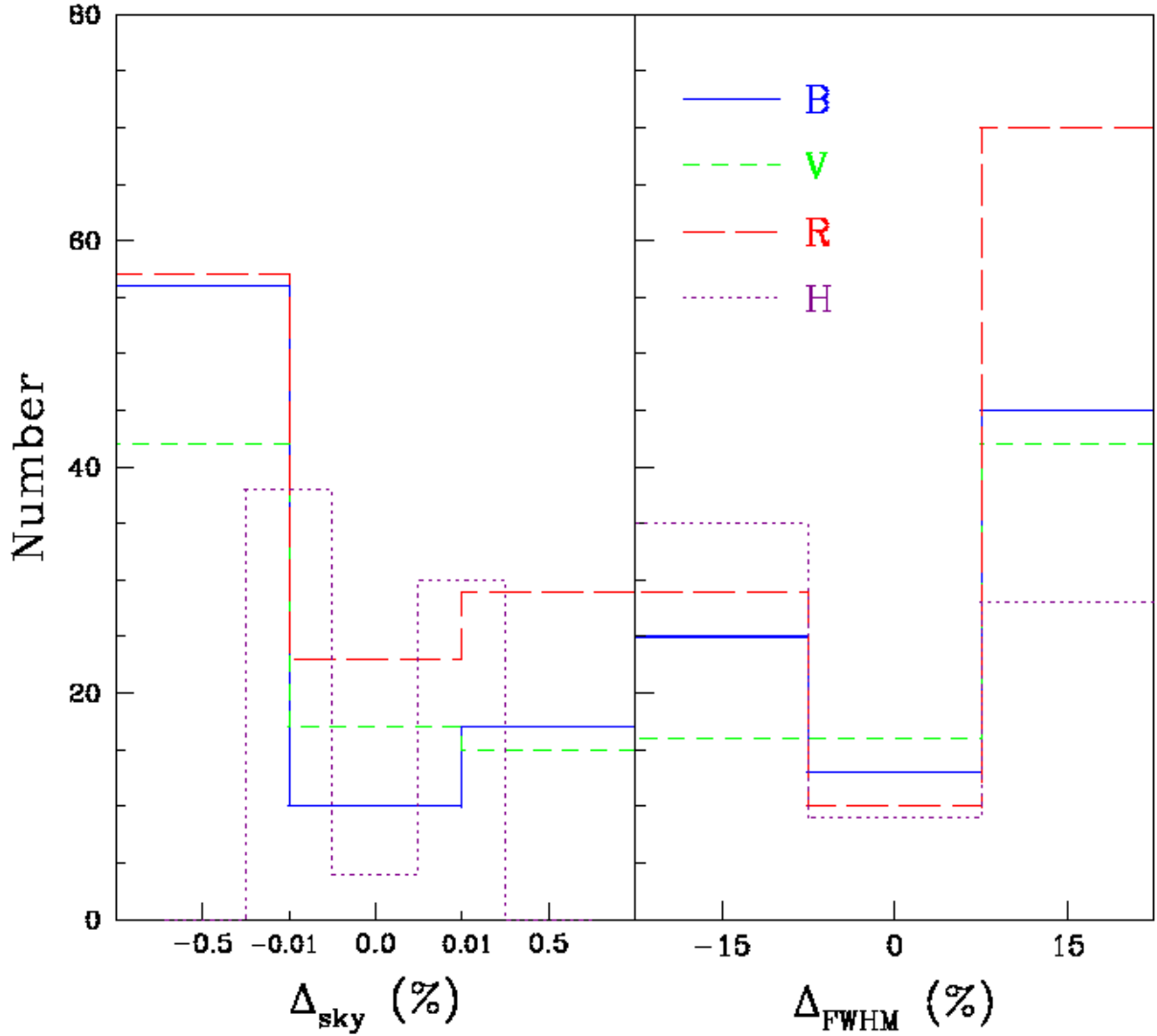}
   \caption{Histograms of sky and seeing FWHM offsets preferred in our
            analysis for all profiles surviving the final cut, separated into
            the four different bands.  Note that the H-band sky error is more
            than an order of magnitude smaller than in the optical (as
            in the actual measurements). \label{fig:skysinghist}}
\end{figure}

\begin{figure}
\plotone{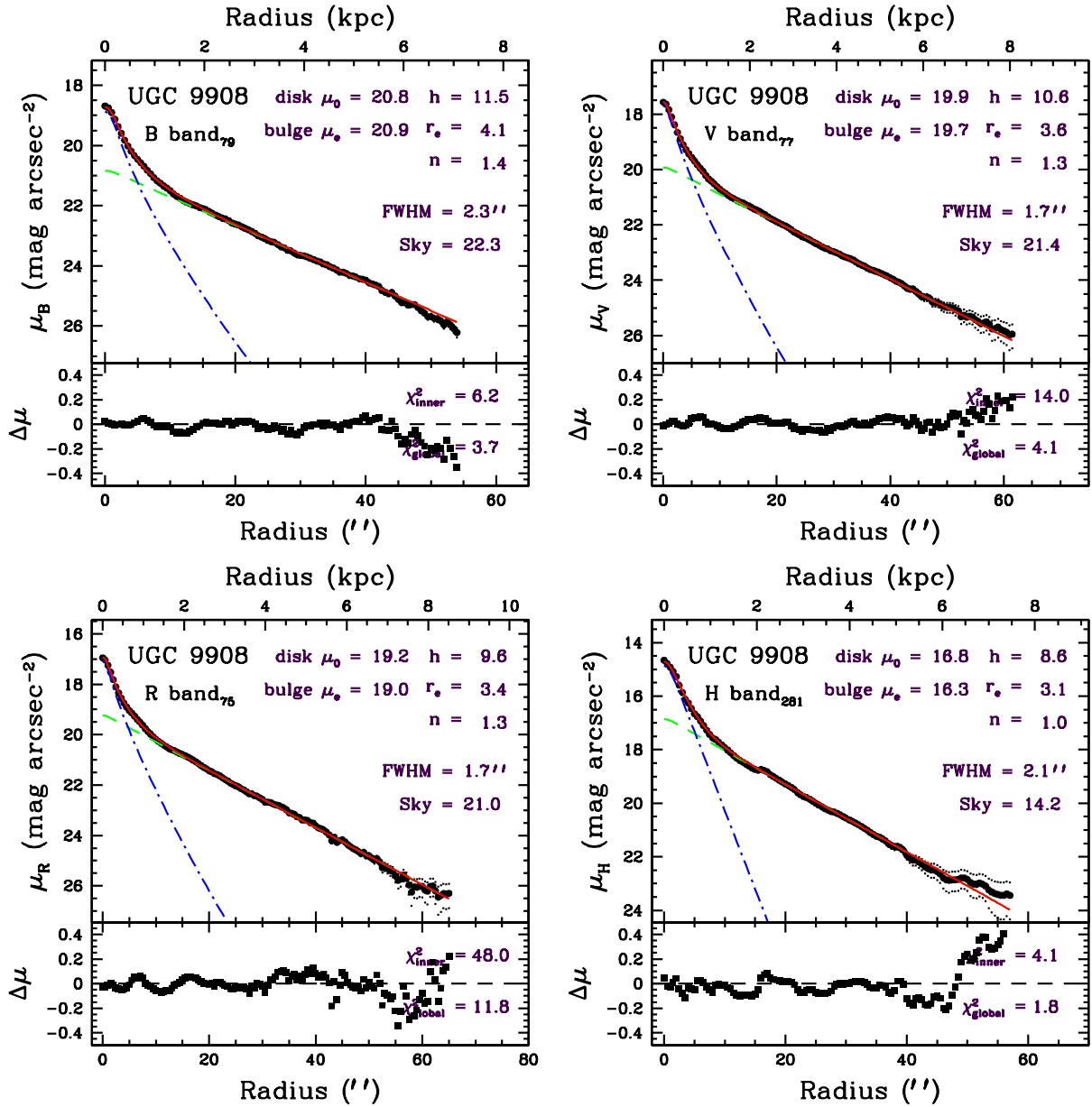}
   \caption{Decomposition results for a Type~I galaxy (UGC 9908).  In the
            upper panels of each plot, the data points and measured sky error
            envelopes are shown with solid black circles and dots respectively.
            The blue dashed-dotted and green
            dashed lines show the bulge and disk fits respectively, and
            the solid red line is the total (bulge+disk) fit.  The fits are all
            seeing-convolved using the best selected seeing values.  The
            bottom panels show the fit residuals where $\Delta \mu(r)
            \equiv data(r)-model(r)$. \label{fig:typeI_dec}}
\end{figure}

\begin{figure}
\plotone{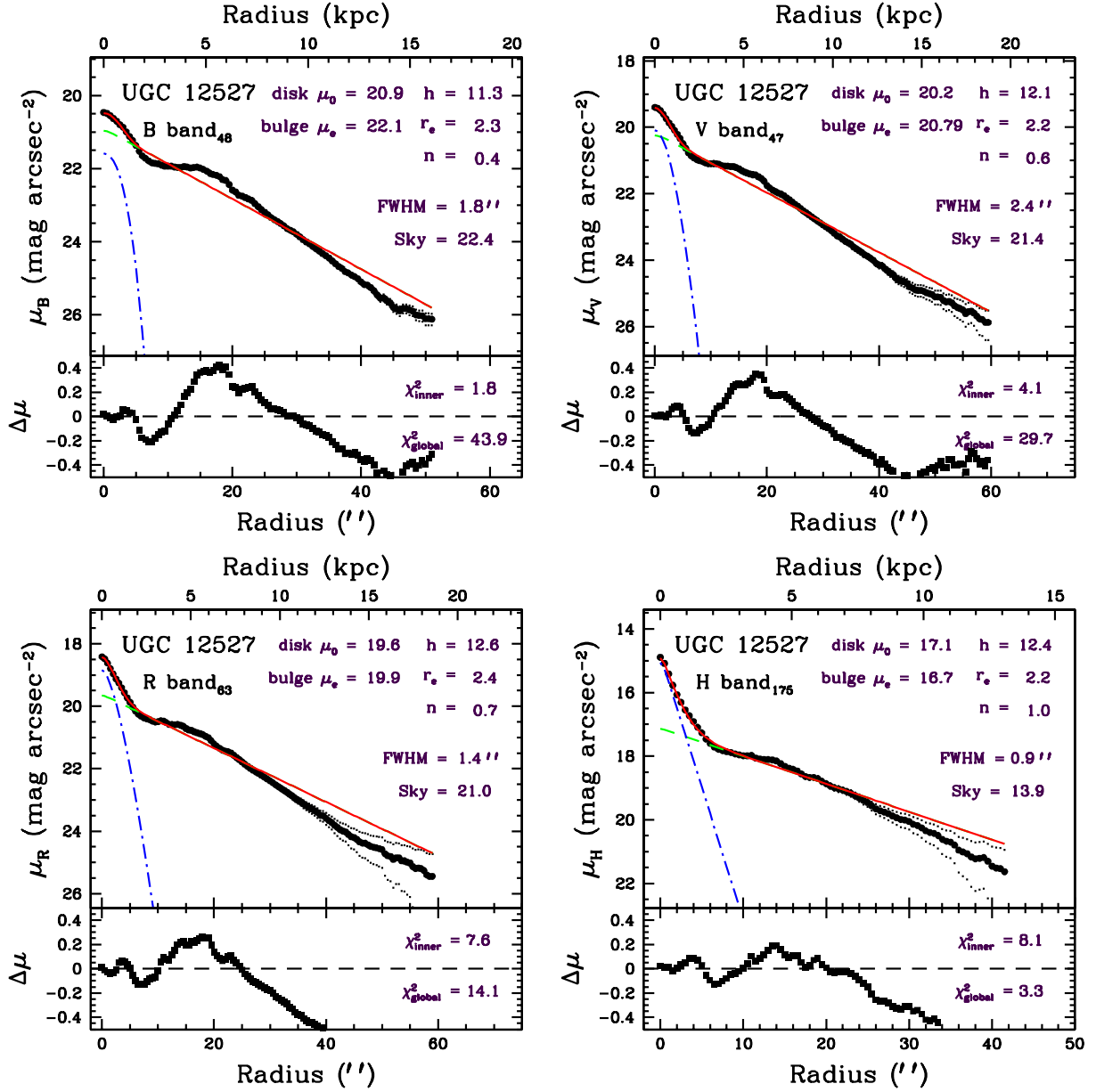}
   \caption{Decomposition results for a Type-II/Transition galaxy (UGC 12527).
           \label{fig:typeII_dec}}
\end{figure}

\begin{figure}
\plotone{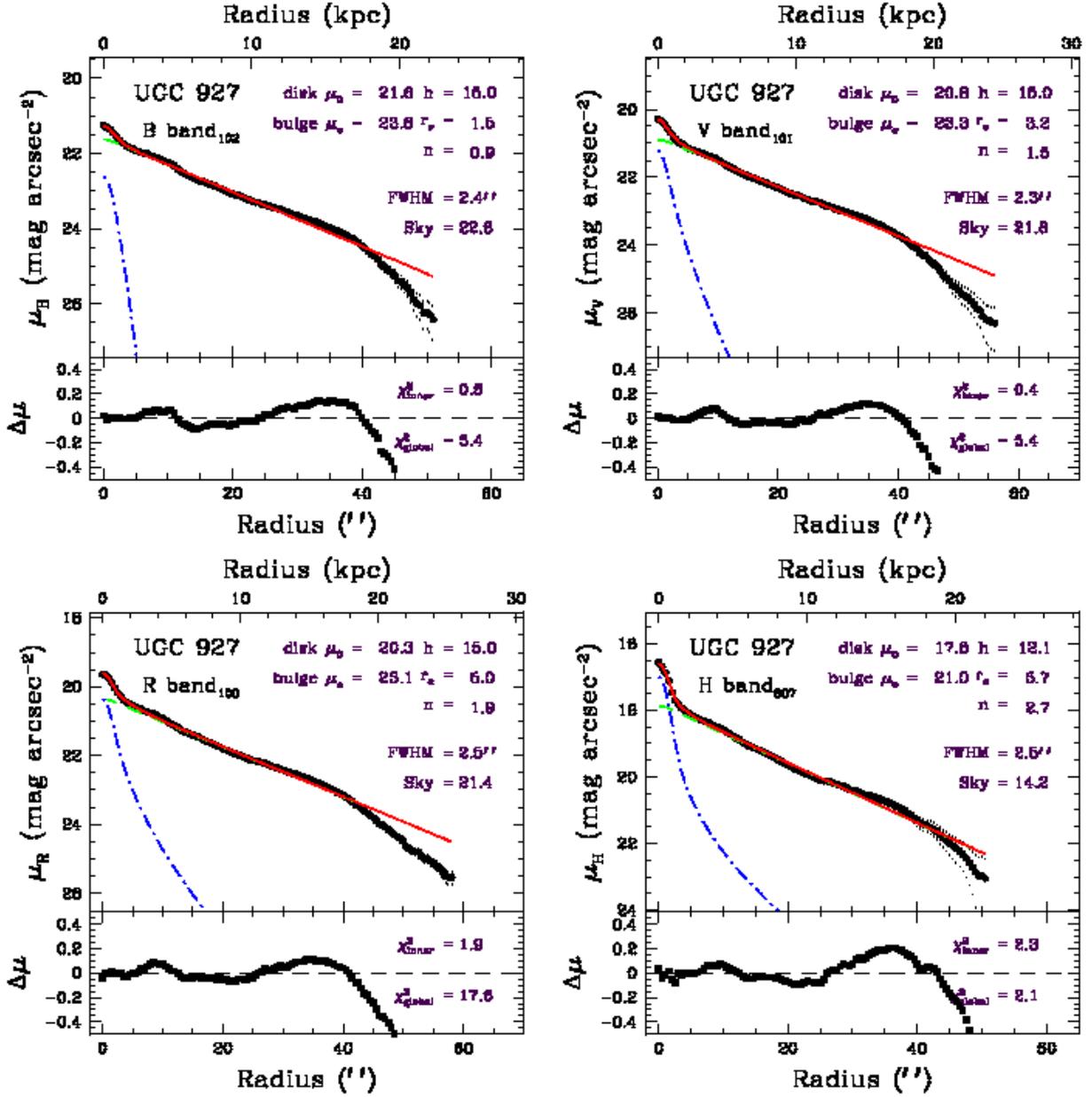}
   \caption{Decomposition results for a galaxy with a truncated disk
           (UGC 927).  Note that sky errors could not account for the 
            truncation. \label{fig:trunc_dec}}
\end{figure}

\begin{figure}
\plotone{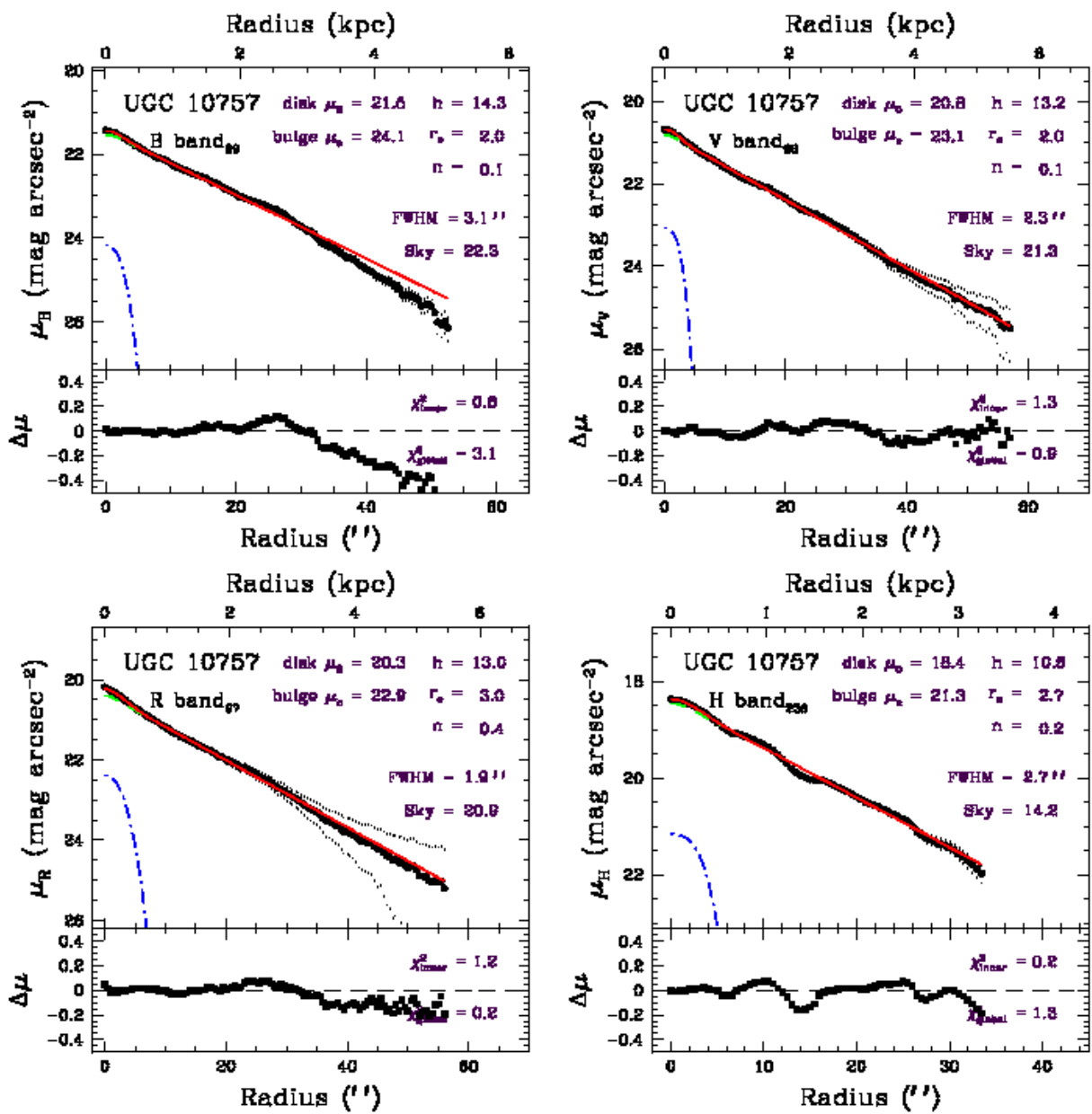}
   \caption{Decomposition results for a galaxy with a ``bulgeless'' disk
           (UGC 10757).
   \label{fig:nobulge_dec}}
\end{figure}

\begin{figure}
\plotone{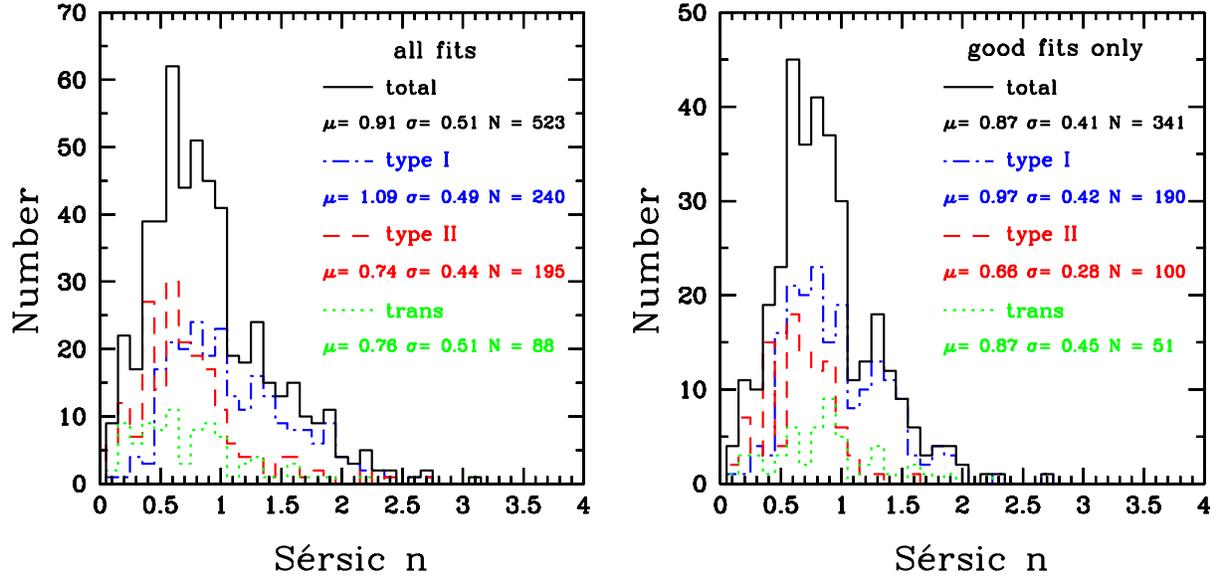}
   \caption{Histograms of \sersic\ $n$ parameter for ``final'' solutions
           (left), and the reduced set of solutions after further visual
            examination (right).  See text for details.
   \label{fig:hist_n_all3}}
\end{figure}

\begin{figure}
\plotone{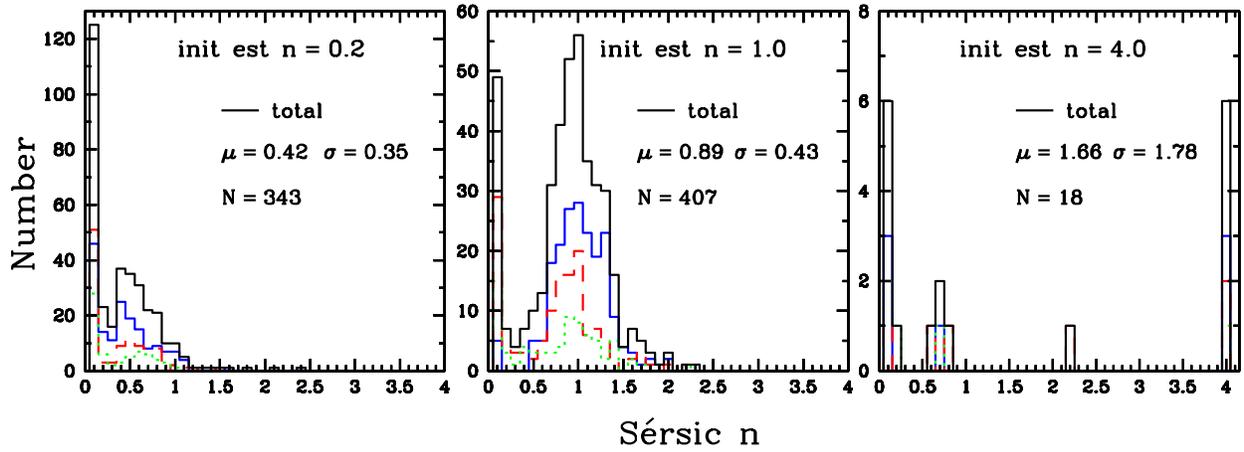}
   \caption{Histograms of \sersic\ $n$ parameter fitting $n$ as a free
    parameter in the decompositions.  Results using three different values for
    the initial estimate of $n$ are shown: $n=0.2$ (left),  $n=1.0$ (middle),
    $n=4.0$ (right).  Note the different y-axis scales in each of the plots.
    The selection criteria for the fits is as described in the text.
    \label{fig:hist_n_floatn}}
\end{figure}

\begin{figure}
\plotone{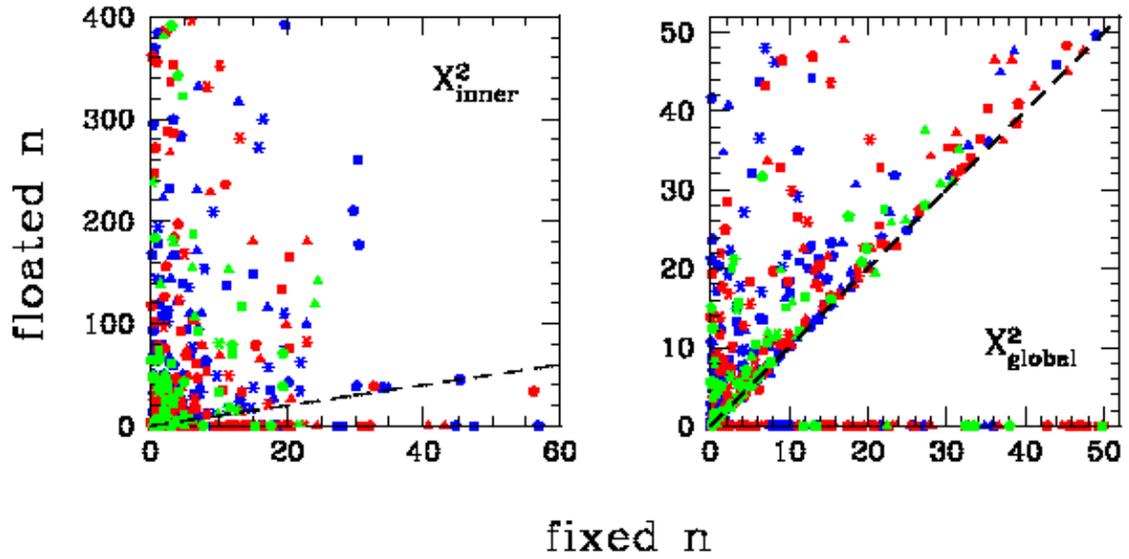}
   \caption{$\chi^2$ comparison of floated $n$ versus fixed $n$ solutions. The
    point types and colors are as follows: B-band (triangles), V-band 
   (squares), R-band (pentagons), H-band (asterisks), Type-I (blue), Type-II 
   (red), and Transition (green).  Note the different axis scales
    for \chisqrin. \label{fig:floatn}}
\end{figure}

\begin{figure}
\plotone{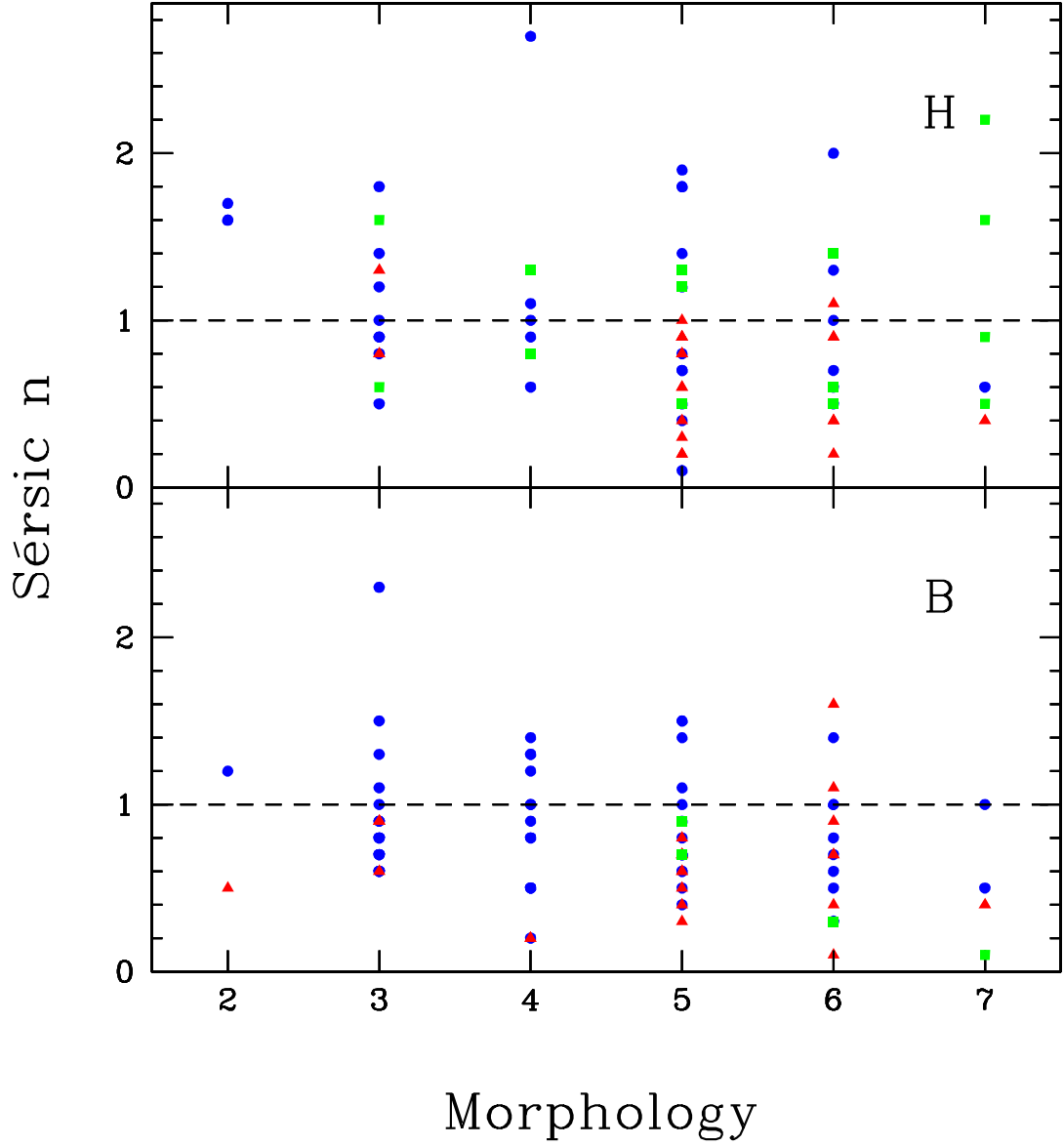}
   \caption{\sersic\ $n$ versus morphological type index.  Blue circles, red 
    triangles, and green squares indicate Type-I, Type-II, and Transition 
    galaxies respectively. \label{fig:sersic_morph}}
\end{figure}

\begin{figure}
\plotone{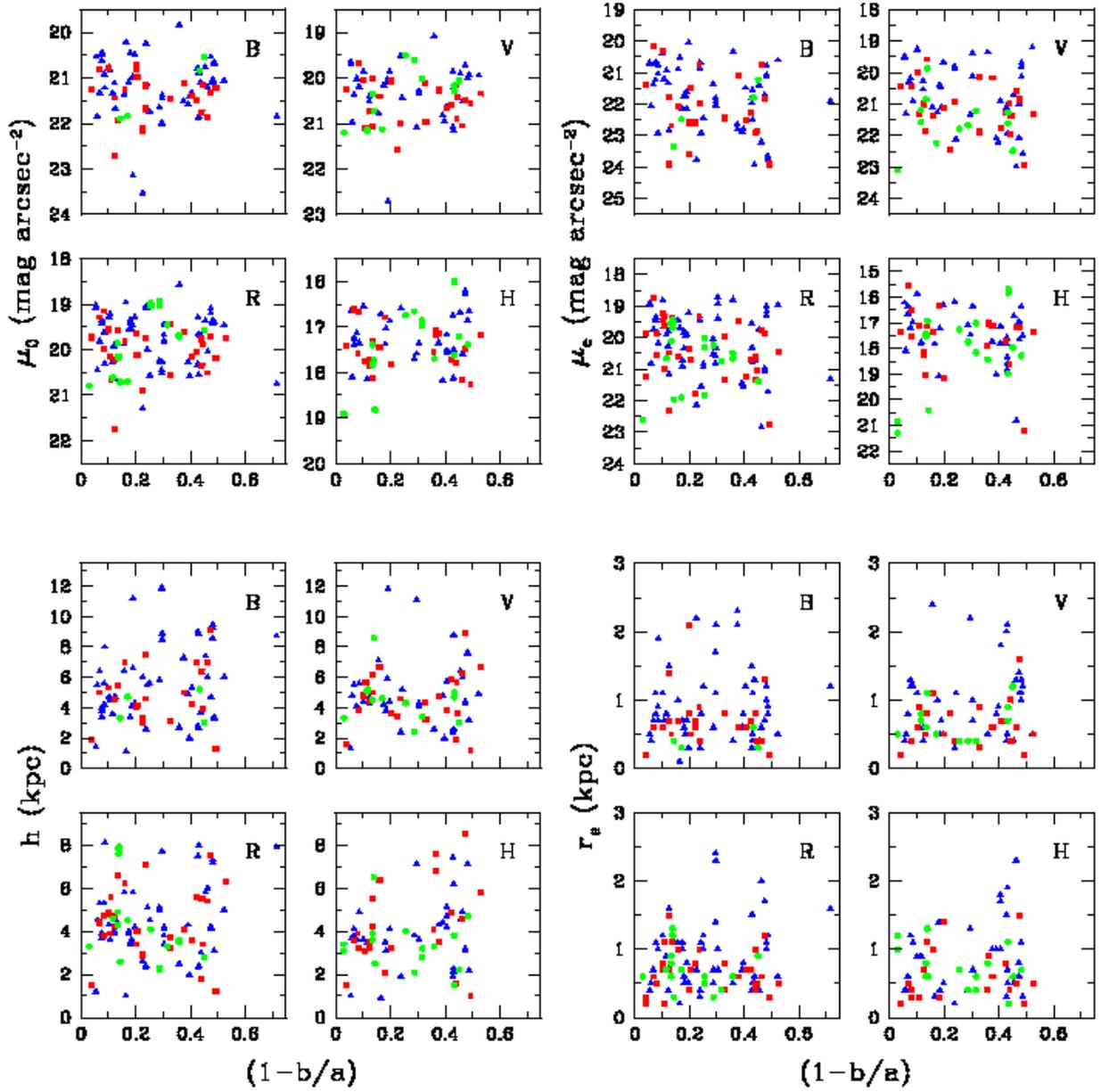}
    \caption{Bulge and disk parameters versus ellipticity $(1-b/a)$.  
    The point types are as follows: Type-I (blue triangles), Type-II 
   (red squares), Transition (green circles). \label{fig:inc_all}}
\end{figure}

\begin{figure}
\plotone{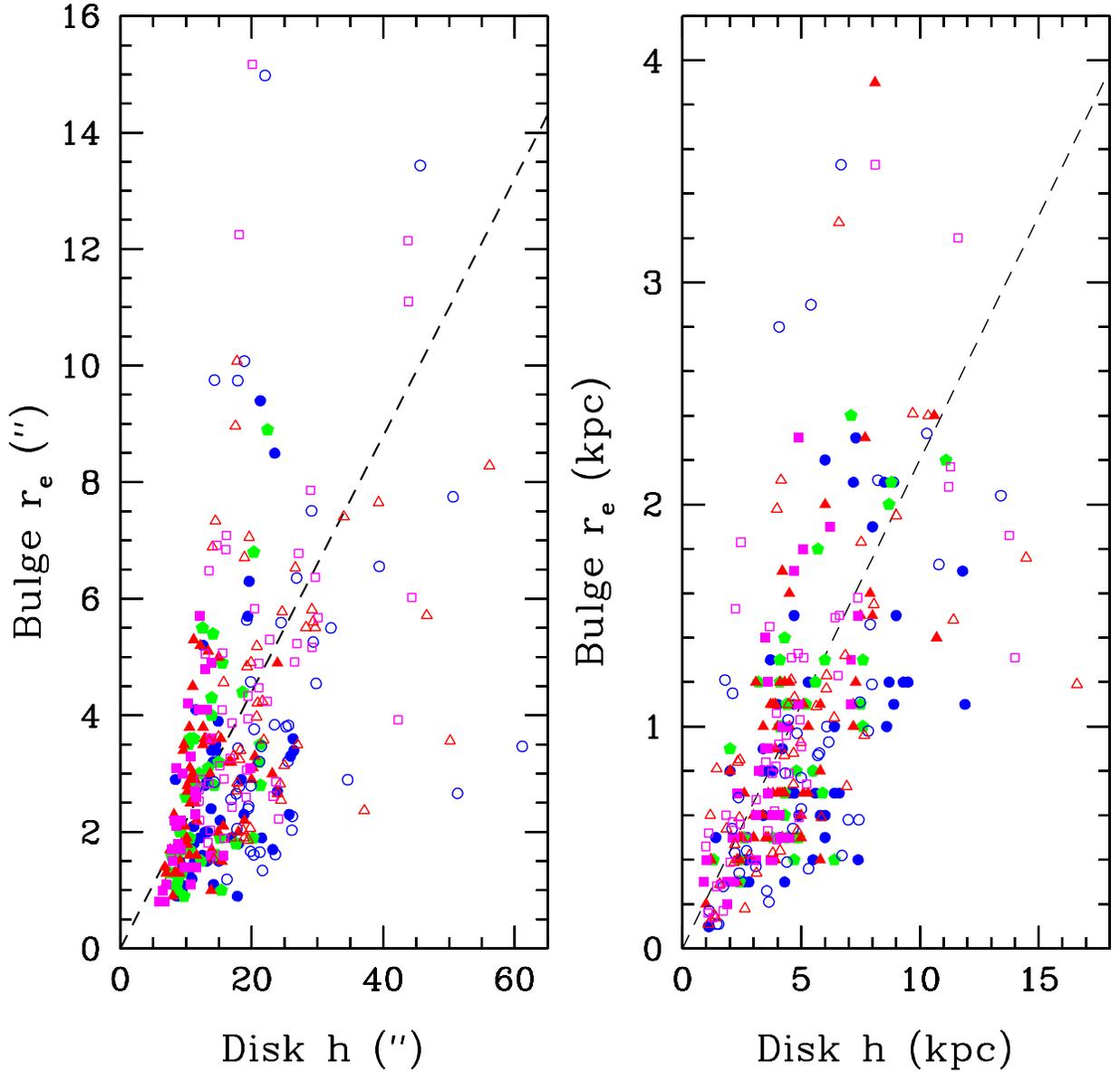}
   \caption{$r_e^{\lambda}$ versus $h^{\lambda}$ for our current Type-I data 
    (solid symbols) and the decompositions of \citet{Graham01} of 
    \citet{dJvdK94}'s data (open symbols).  Blue circles are B-band, green
    pentagons (our data only) are V-band, red triangles are R-band, and 
    magenta squares are H-band (us) and K-band (\citet{Graham01}).  The 
    dashed lines have
    a slope $\langle r_e/h \rangle = 0.22$ for late-type spirals.  Note 
    that the large dispersions in the $r_e^\lambda$ and $h^\lambda$ 
    (Table~\ref{tab:means}) counteract to yield significant $r_e/h$ 
    correlations.  The left plot is in apparent units (arcsec) and the right 
    plot shows the physical scale in kpc.  The discrete nature of our data 
    in the right plot is due to the limited precision of the $r_e^{\lambda}$ 
    measurement (one decimal).
    \label{fig:rb_rd}}
\end{figure}

\begin{figure}
\plotone{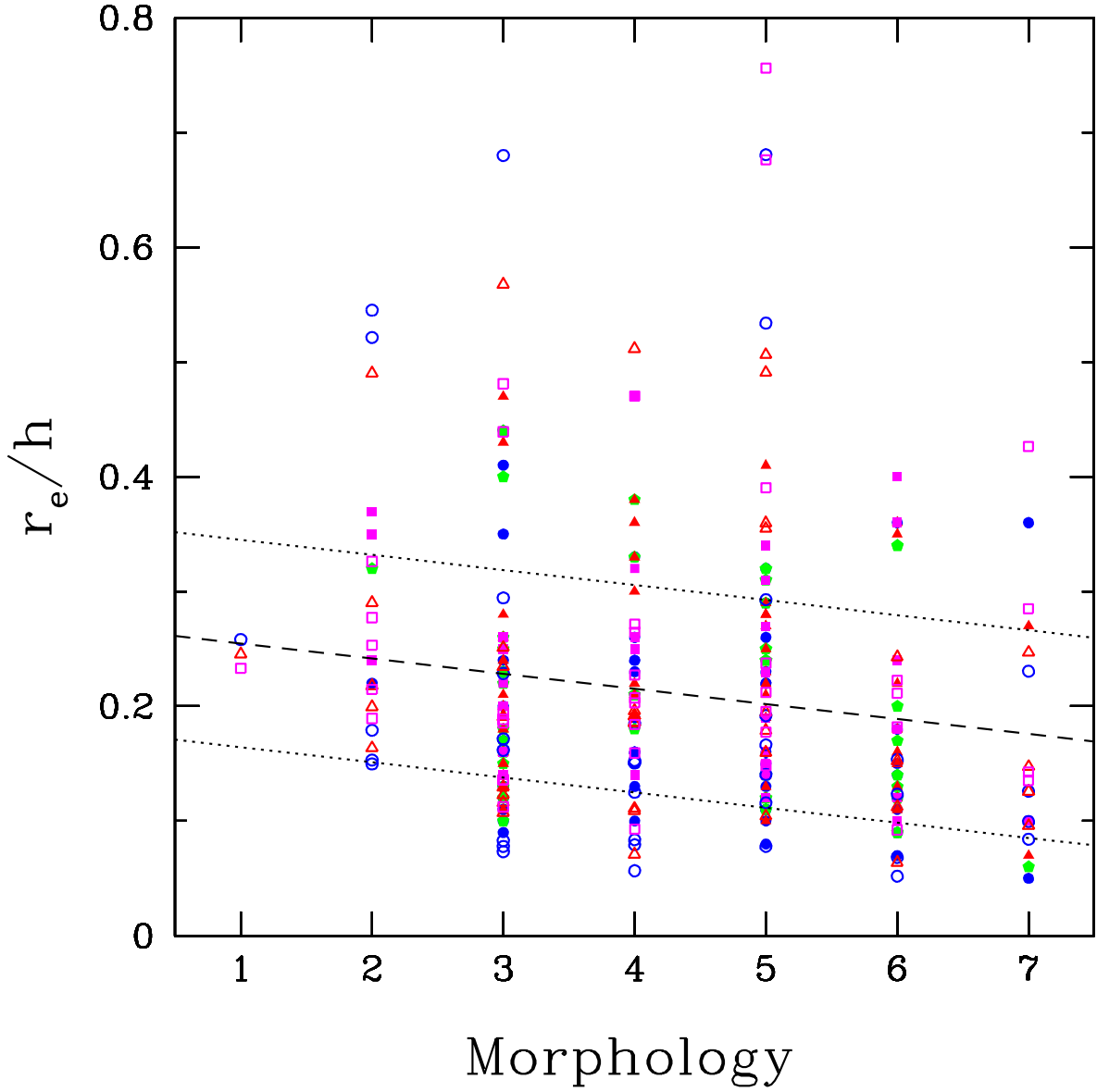}
    \caption{Distribution of $r_e/h$ with Hubble types for our Type-I galaxies
     and those of Graham (2001).  Symbols and colors are as in 
     Fig.~\ref{fig:rb_rd}.  The dashed line describes the fit
     $\langle r_e/h \rangle = 0.20- 0.013(T-5)$ with 1$\sigma=0.09$ errors
     (dotted lines) based on our data only. 
     \label{fig:reh_vs_morph}}
\end{figure}

\begin{figure}
\plotone{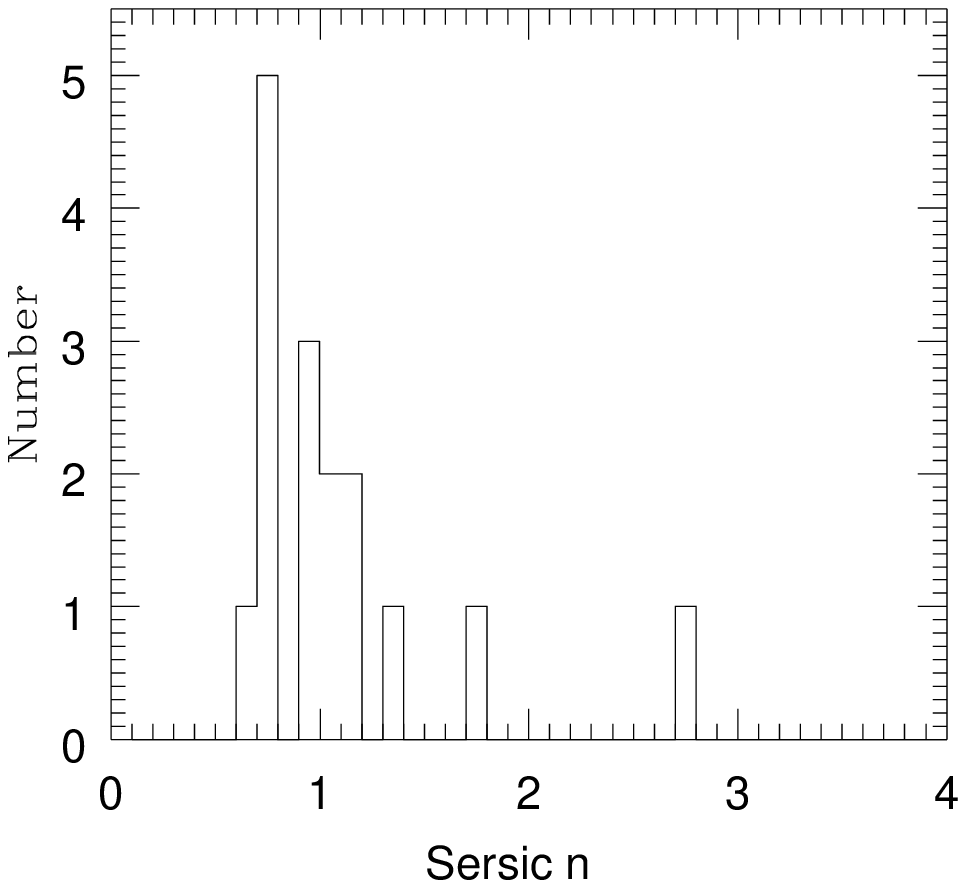}
   \caption{Distribution of the \sersic\ $n$ parameter
            from cosmological simulations by
            \citet{ScaTissera02}.
   \label{fig:tissera}}
\end{figure}

\begin{figure}
\plotone{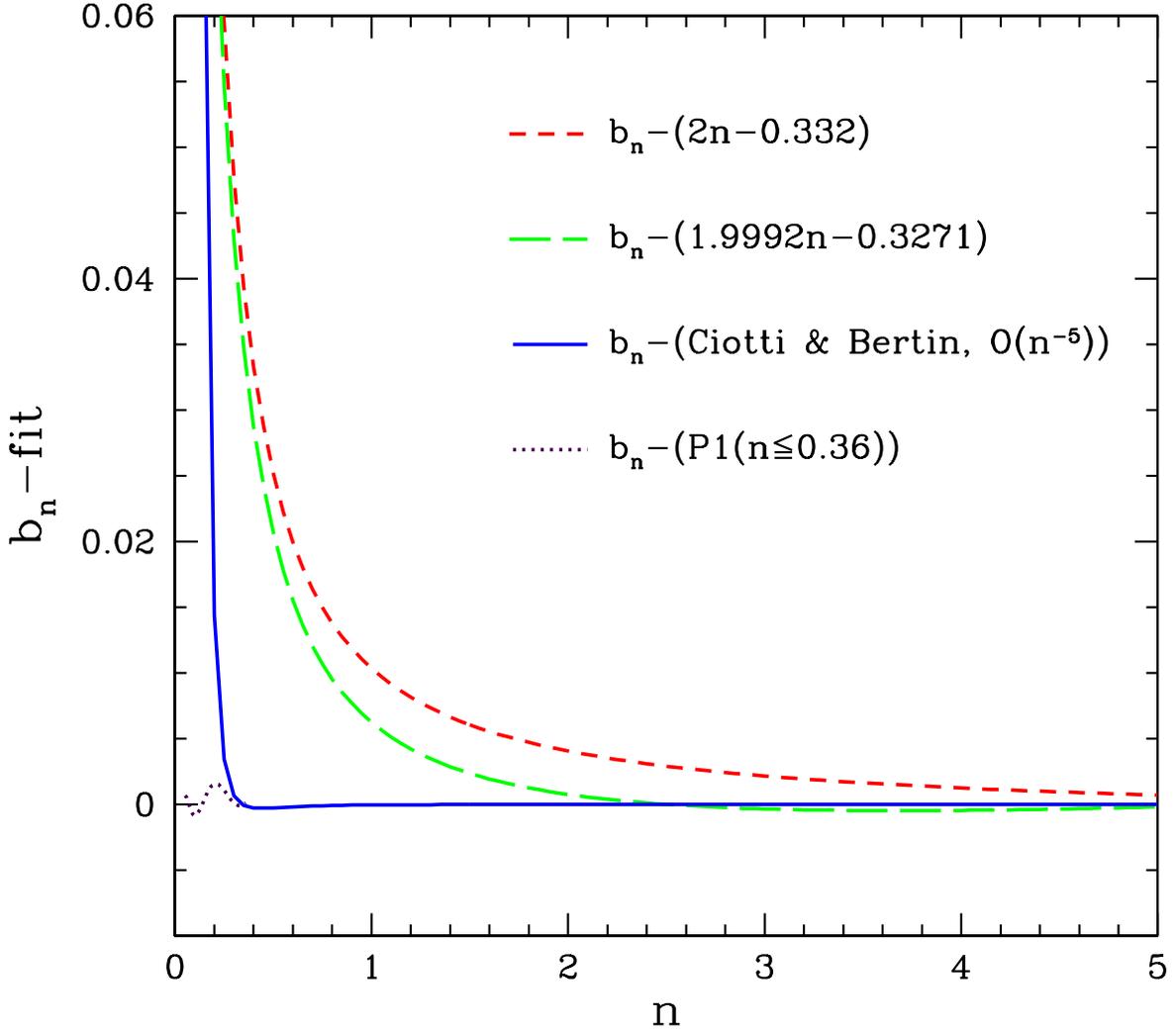}
   \caption{Difference between the 
     exact numerical value for $b_n$ and several commonly adopted 
     approximations.  The short (red) and long (green) dashed lines are the 
     two most commonly used approximations found in the literature.  The 
     solid blue line shows Ciotti \& Bertin's asympotic expansion and the 
     dotted purple line depicts our adopted extension at $n \leq 0.36$.
     \label{fig:bncomp}}
\end{figure}

{\it Column (3)}: Sky brightness in \magarc, measured from 4 sky boxes
located between the detector edges and a fair distance away from the galaxy.
Typical rms sky errors, computed from the deviations of the mean sky counts
in those sky boxes, are $\sim 0.5-1.0$\% in the optical and ~0.05\% in
the IR.  The subscripts indicate the sky offset preferred by our selection
process as described in \S~\ref{subsubsec:singskytreat} and
\S~\ref{subsec:filtering} (and see Fig.~\ref{fig:skysinghist}), where
``$+$'' and ``$-$'' indicate 0.5\% for optical and 0.01\% for H-band
over- and under-subtracted skies respectively.  No subscript indicates that
the measured sky was preferred;

{\it Column (4)}: Seeing FWHM values, computed as the average of the FWHMs
of all non-saturated stars measured automatically on each image frame;
typically 10 to 30 measurements per image were used for each FWHM estimate.
The accuracy of the seeing estimate per image is roughly 20\% for the
optical bands and 30\% for the H-band.  The subscripts indicate the seeing
offset preferred by our selection process as described in
\S~\ref{subsubsec:singskytreat} and \S~\ref{subsec:filtering}, (and see
Fig.~\ref{fig:skysinghist}) where ``$+$'' and ``$-$'' indicate 15\%
over- and under-estimated seeing FWHM respectively.
No subscript indicates that the measured seeing FWHM was preferred;

The upper and lower boundaries in the remaining columns correspond to the
maximum and minimum values of the $\leq50$ (out of 1080 total) solutions
left after filtering (see \S~\ref{subsec:filtering});

{\it Column (5)}: Best fit \sersic\ $n$ bulge shape parameter;

{\it Column (6)}: Bulge effective surface brightness, $\mu_e$, in \magarc,
corrected for Galactic extinction and cosmological redshift
dimming, as described in \S~\ref{sec:sbcorr};

{\it Column (7)}: Bulge effective radius, $r_e$, in arcseconds;

{\it Column (8)}: Bulge effective radius, $r_e$, in kpc.  Converted to
a physical scale using the Local Standard of Rest velocity, $V_{LG}$
(see Paper II);

{\it Column (9)}: Exponential disk central surface brightness,
$\mu_0$ in \magarc, corrected for Galactic extinction and
cosmological redshift dimming as described in
\S~\ref{sec:sbcorr};

{\it Column (10)}: Exponential disk scale length $h$, in arcseconds;

{\it Column (11)}: Exponential disk scale length $h$, in kpc.
Converted to a physical scale using the Local Standard of Rest
velocity, $V_{LG}$ (see Paper II);

{\it Column (12)}: Bulge-to-disk luminosity ratio, $B/D$, calculated using
Eq.~\ref{eq:bdrat} in \S~\ref{subsec:algorithm}.

\clearpage

\clearpage

\clearpage

\clearpage

\clearpage

\clearpage

\clearpage

\clearpage

\clearpage

\clearpage

\clearpage

\clearpage

\clearpage

\clearpage 

\clearpage

\clearpage

\clearpage

\clearpage

\clearpage

\clearpage

\clearpage

\clearpage

\clearpage

\clearpage

\begin{deluxetable}{lc|cccccccccc|c}
\tabletypesize{\normalsize}
\tablewidth{0pt}
\tablecaption{Table of mean values and mean rms deviations for repeat
observations. \label{tab:rms}
}\tablehead{
\multicolumn{2}{c|}{} &
\colhead{$\overline{n}$} &
\colhead{$\overline{\sigma_n}$} &
\colhead{${\overline{\mu_{e}}}$} &
\colhead{$\overline{{\sigma_{\mu_{e}}}}$} &
\colhead{$\overline{r_e}$} &
\colhead{$\overline{\sigma_{r_e}}$} &
\colhead{$\overline{\mu_{0}}$} &
\colhead{$\overline{\sigma_{\mu_{0}}}$} &
\colhead{$\overline{h}$} &
\colhead{$\overline{\sigma_{h}}$} &
\multicolumn{1}{|c}{$N$} \\

\multicolumn{2}{c|}{} &
\colhead{} &
\colhead{} &
\colhead{\scriptsize{(mag}/$\sq^{\arcsec}$)} &
\colhead{} &
\colhead{\scriptsize{(kpc)}} &
\colhead{} &
\colhead{\scriptsize{(mag}/$\sq^{\arcsec}$)} &
\colhead{} &
\colhead{\scriptsize{(kpc)}} &
\colhead{} &
\multicolumn{1}{|c}{} }
\startdata
I  & B & 1.00 & 0.13 & 21.88 & 0.19 & 1.13 & 0.08 & 20.90 & 0.03 & 5.35 & 0.18 & 7 \\
   & V & 0.98 & 0.07 & 20.16 & 0.27 & 0.81 & 0.11 & 19.96 & 0.03 & 4.06 & 0.08 & 7 \\
   & R & 0.88 & 0.16 & 20.02 & 0.17 & 0.77 & 0.11 & 19.59 & 0.04 & 3.76 & 0.07 & 13 \\
   & H & 0.95 & 0.14 & 17.52 & 0.26 & 0.85 & 0.14 & 17.33 & 0.10 & 3.49 & 0.17 & 5 \\
\hline
   & \hidewidth{Total} & 0.94 & 0.13 & 20.07 & 0.21 & 0.87 & 0.11 & 19.61 & 0.05 & 4.13 & 0.11 & 32 \\
\hline
\hline
II & B & 0.72 & 0.17 & 22.84 & 0.28 & 1.05 & 0.36 & 21.45 & 0.06 & 3.60 & 0.13 & 2 \\
   & R & 0.61 & 0.19 & 20.00 & 0.34 & 0.41 & 0.05 & 19.80 & 0.02 & 3.01 & 0.02 & 4 \\
   & H & 0.68 & 0.25 & 17.61 & 0.07 & 0.80 & 0.00 & 17.45 & 0.09 & 5.28 & 0.39 & 2 \\
\enddata
\end{deluxetable}

\begin{deluxetable}{c|cccccrr}
\tabletypesize{\normalsize}
\tablewidth{0pt}
\tablecaption{Comparison of B/D decomposition parameters.
\label{tab:us_vs_dJ}}
\tablehead{
\multicolumn{1}{c|}{Author} &
\colhead{UGC}  &
\colhead{band} &
\colhead{${\mu_{0}}$} &
\colhead{$h$}  &
\colhead{$\mu_{e}$} &
\colhead{$r_e$}  &
\colhead{$n$}  \\
\multicolumn{1}{c|}{} &
\colhead{} &
\colhead{} &
\colhead{(mag/$\sq^{\arcsec}$)} &
\colhead{(\arcsec)} &
\colhead{(mag/$\sq^{\arcsec}$)} &
\colhead{(\arcsec)}  &
\colhead{}  }
\startdata
Graham  &   463 & B & 20.77 & 13.0 & 20.51 & 1.6 & 0.4 \\
de~Jong &   463 & B & 20.76 & 13.5 & 20.60 & 1.3 & 1.0 \\
us      &   463 & B & 20.59 & 14.2 & 20.15 & 1.1 & 0.4 \\ \hline
de~Jong &   463 & H & 16.80 & 12.0 & 16.73 & 1.9 & 1.0 \\
us      &   463 & H & 17.11 & 13.7 & 16.56 & 1.6 & 0.5 \\ \hline
Graham  &   3080 & B & 22.20 & 19.5 & 22.42 & 1.4 & 0.3 \\
de~Jong &   3080 & B & 21.99 & 17.2 & 19.88 & 0.2 & 1.0 \\
us      &   3080 & B & 21.99 & 24.3 & 22.74 & 2.0 & 0.5 \\
\enddata
\tablecomments{Parameter comparison between \citet{deJong96a},
\citet{Graham01}, and
us.  Graham's decompositions (B-band only as Graham did not re-analyse
de~Jong's H-band data) use de~Jong's 1D brightness profiles.  Our
decompositions are based upon data discussed in this paper and presented
in Paper II.  The surface
brightnesses have not been corrected for Galactic or internal extinction in
order to compare with \citet{deJong96a} and \citet{Graham01}.
Repeat observations
are listed for UGC 3140 B-band.  The galaxies UGC 463 and 3080 are Type-II
systems; UGC 3140 is a Type-I galaxy.  See Graham (2001; Fig. 6) for examples
of his decompositions (esp. for UGC 463).}
\end{deluxetable}

\begin{deluxetable}{ccccccc}
\tabletypesize{\normalsize}
\tablewidth{0pt}
\tablecaption{Mean and standard deviation, $\mu(\sigma)$, for bulge and disk parameters
for different types and bandpasses.  N denotes the number of data
points.  Multiple observations count as independent entries.
\label{tab:means}}
\tablehead{
\colhead{Param} &
\colhead{Type} &
\colhead{} &
\colhead{B} &
\colhead{V} &
\colhead{R} &
\colhead{H} }
\startdata

 &  I & N & 56 & 39 & 59 & 34 \\ [-0pt]
 & II & N & 23 & 23 & 31 & 22 \\ [-0pt]
 & Tr & N &  4 & 12 & 19 & 16 \\ [+8pt]

 $n$  & I  & $\mu(\sigma)$ & 0.88 (0.38) & 1.02 (0.42) & 0.95 (0.35) & 1.11 (0.57) \\ [-0pt]
      & II & $\mu(\sigma)$ & 0.63 (0.33) & 0.65 (0.25) & 0.67 (0.23) & 0.68 (0.32) \\ [-0pt]
      & Tr & $\mu(\sigma)$ & 0.50 (0.37) & 0.89 (0.41) & 0.79 (0.40) & 1.04 (0.50) \\ [+8pt]

$r_e$ & I  & $\mu(\sigma)$ & 1.00 (0.77) & 1.07 (0.81) & 0.96 (0.62) & 0.87 (0.52) \\ [-0pt]
(kpc) & II & $\mu(\sigma)$ & 0.72 (0.41) & 0.65 (0.32) & 0.66 (0.32) & 0.65 (0.36) \\
\enddata
\end{deluxetable}

\begin{deluxetable}{rcllrrrrrrrr}
\tabletypesize{\scriptsize}
\tablecaption{Decomposition Results for Type I Profiles\label{tab:decomps}}
\tablewidth{0pt}
\tablehead{
\colhead{Profile} &
\colhead{$\varepsilon$} &
\colhead{Sky} &
\colhead{\hidewidth{\tiny{FWHM}}} & 
\colhead{$n$} & 
\colhead{${\mu_{e}}$} & 
\colhead{$r_e$} & 
\colhead{$r_e$} & 
\colhead{$\mu_{0}$} & 
\colhead{$h$} & 
\colhead{$h$} & 
\colhead{$B/D$} 
\\ 

\colhead{\hidewidth{\tiny{UGC obs band}}}& 
\colhead{\hidewidth{\tiny{(1-$b/a$)}}} & 
\colhead{\hidewidth{\tiny{\,\,(mag/$\sq^{\arcsec}$)}}} & 
\colhead{(\arcsec)} & 
\colhead{} & 
\colhead{(\tiny{mag}/$\sq^{\arcsec}$)} & 
\colhead{(\arcsec)} & 
\colhead{(kpc)} & 
\colhead{(\tiny{mag}/$\sq^{\arcsec}$)} & 
\colhead{(\arcsec)} & 
\colhead{(kpc)} & 
\colhead{} \\ 
\multicolumn{1}{c}{(1)} & \multicolumn{1}{c}{(2)} & \multicolumn{1}{c}{(3)} & \multicolumn{1}{c}{(4)} & \multicolumn{1}{c}{(5)} & \multicolumn{1}{c}{(6)} & \multicolumn{1}{c}{(7)} & \multicolumn{1}{c}{(8)} & \multicolumn{1}{c}{(9)} & \multicolumn{1}{c}{(10)} & \multicolumn{1}{c}{(11)} & \multicolumn{1}{c}{(12)}}
\startdata
   10  044  B & 0.295 & $22.33_{-}$& $ 1.8_{-}$& $1.0_{-0.0}^{+0.5}$& $21.92_{- 0.00}^{+ 0.59}$& $  2.2_{-  0.0}^{+  0.6}$& $  1.7_{-  0.0}^{+  0.5}$& $22.01_{- 0.00}^{+ 0.08}$& $ 15.2_{-  0.0}^{+  1.1}$& $ 11.8_{-  0.0}^{+  0.9}$& $ 0.04_{- 0.00}^{+ 0.01}$\\ [2pt]
   10  043  V & 0.295 & $21.55_{-}$& $ 2.0_{ }$& $1.4_{-0.0}^{+0.5}$& $21.32_{- 0.00}^{+ 0.77}$& $  2.8_{-  0.0}^{+  1.6}$& $  2.2_{-  0.0}^{+  1.3}$& $21.09_{- 0.00}^{+ 0.19}$& $ 14.3_{-  0.0}^{+  1.6}$& $ 11.1_{-  0.0}^{+  1.2}$& $ 0.07_{- 0.00}^{+ 0.03}$\\ [2pt]
   10  042  R & 0.295 & $21.09_{-}$& $ 1.8_{ }$& $1.5_{-0.0}^{+0.5}$& $20.82_{- 0.00}^{+ 0.80}$& $  3.0_{-  0.0}^{+  2.1}$& $  2.4_{-  0.0}^{+  1.7}$& $20.50_{- 0.00}^{+ 0.22}$& $ 13.7_{-  0.0}^{+  1.6}$& $ 10.6_{-  0.0}^{+  1.2}$& $ 0.08_{- 0.00}^{+ 0.04}$\\ [2pt]
   10  754  H & 0.295 & $13.96_{-}$& $ 1.2_{ }$& $1.3_{-0.1}^{+0.3}$& $17.48_{- 0.17}^{+ 0.46}$& $  1.7_{-  0.1}^{+  0.5}$& $  1.3_{-  0.1}^{+  0.4}$& $17.58_{- 0.04}^{+ 0.10}$& $  9.2_{-  0.0}^{+  0.7}$& $  7.1_{-  0.0}^{+  0.5}$& $ 0.08_{- 0.00}^{+ 0.01}$\\ [2pt]
   16  036  B & 0.157 & $22.25_{+}$& $ 1.4_{-}$& $1.2_{-0.1}^{+0.1}$& $20.90_{- 0.15}^{+ 0.09}$& $  2.9_{-  0.2}^{+  0.0}$& $  1.0_{-  0.1}^{+  0.0}$& $21.36_{- 0.03}^{+ 0.00}$& $ 18.4_{-  0.5}^{+  0.0}$& $  6.4_{-  0.2}^{+  0.0}$& $ 0.08_{- 0.01}^{+ 0.00}$\\ [2pt]
   16  039  V & 0.157 & $21.44_{+}$& $ 2.1_{ }$& $1.9_{-0.1}^{+0.1}$& $21.08_{- 0.15}^{+ 0.15}$& $  6.8_{-  0.6}^{+  0.6}$& $  2.4_{-  0.2}^{+  0.2}$& $20.78_{- 0.06}^{+ 0.03}$& $ 20.3_{-  0.7}^{+  0.1}$& $  7.1_{-  0.2}^{+  0.0}$& $ 0.22_{- 0.02}^{+ 0.02}$\\ [2pt]
   16  034  R & 0.157 & $20.53_{ }$& $ 1.4_{-}$& $1.4_{-0.2}^{+0.0}$& $19.45_{- 0.31}^{+ 0.00}$& $  3.2_{-  0.3}^{+  0.0}$& $  1.1_{-  0.1}^{+  0.0}$& $19.88_{- 0.07}^{+ 0.00}$& $ 16.5_{-  0.8}^{+  0.0}$& $  5.8_{-  0.3}^{+  0.0}$& $ 0.12_{- 0.00}^{+ 0.01}$\\
\enddata
\end{deluxetable}

\end{document}